\documentclass[12pt,twoside]{article}   %For printing on two sides of page
\usepackage[super,sort&compress,comma]{natbib}
\usepackage{fancyhdr}		%Gives headers and footers defined below
\newcommand{\captionv}[3]{\begin{center}\parbox{#1cm}{\caption[#2]{{\sf #3}}}
        \end{center}}
\usepackage[section]{placeins}   %
\usepackage{graphicx}

\makeatletter \renewcommand\@biblabel[1]{$^{#1}$} \makeatother
\newcommand{\cen}[1]{\begin{center} #1 \end{center}}

\usepackage[mathlines]{lineno}
\usepackage{hyperref}
\hypersetup{ colorlinks,
    citecolor=blue,
    filecolor=blue,
    linkcolor=blue,
    urlcolor=blue
}

\usepackage{xcolor}
\definecolor{gray}{rgb}{0.6,0.6,0.6}
\definecolor{red}{rgb}{0.85,0,0}
\definecolor{green}{rgb}{0,0.85,0}
\definecolor{blue}{rgb}{0,0,0.85}
\definecolor{beige}{rgb}{0.92,0.87,0.78}
\usepackage[all]{hypcap}    %causes link to figures to go to figure, not caption
\usepackage{amsmath}
\usepackage[linesnumbered,ruled,vlined]{algorithm2e}
\usepackage{adjustbox}
\usepackage{threeparttable}
\usepackage{multirow}
\usepackage{arydshln}

\colorlet{yellow}{black}
\colorlet{purple}{black}
\colorlet{cyan}{black}
\colorlet{red}{black}

\begin{document}

\cen{\sf {\Large {\bfseries Automatic view plane prescription for cardiac magnetic resonance imaging via supervision by spatial relationship between views} \\
\vspace*{10mm}
Dong Wei, Yawen Huang, Donghuan Lu, Yuexiang Li, and\\Yefeng Zheng} \vspace{5mm}\\
Tencent Jarvis Lab, Shenzhen, 518057, China
\vspace{5mm}\\
}

\pagenumbering{roman}
\setcounter{page}{1}
\pagestyle{plain}
\textbf{Correspondence:} Yefeng Zheng (yefengzheng@tencent.com) \\
% note, probably best not to use a student's e-mail as it won't be valid for
% very long.

\begin{abstract}
\noindent {\bf Background:} View planning for the acquisition of cardiac magnetic resonance (CMR) imaging remains a demanding task in clinical practice.\\
{\bf Purpose:} Existing approaches to its automation relied either on an additional volumetric image not typically acquired in clinic routine, or on laborious manual annotations of cardiac structural landmarks.
This work presents a clinic-compatible, annotation-free system for automatic CMR view planning.\\
{\bf Methods:} The system mines the spatial relationship---more specifically, locates the intersecting lines—between the target planes and source views, and trains U-Net-based deep networks to regress heatmaps defined by distances from the intersecting lines.
%by the intersecting lines.
On the one hand, the intersection lines {\color{red}are} the prescription lines prescribed by the technologists at the time of image acquisition using cardiac landmarks, and retrospectively identified from the spatial relationship.
On the other hand, as the spatial relationship is self-contained in properly stored data, e.g., in the DICOM format, the need for additional manual annotation is eliminated.
In addition, the interplay of the multiple target planes predicted in a source view is utilized in a stacked hourglass architecture consisting of repeated U-Net-style building blocks to gradually improve the regression.
Then, a multi-view planning strategy is proposed to aggregate information from the predicted heatmaps for all the source views of a target plane, for a globally optimal prescription, mimicking the similar strategy practiced by skilled human prescribers.
For performance evaluation, the retrospectively identified planes prescribed by the technologists are used as the ground truth, and the plane angle differences and localization distances between the planes prescribed by our system and the ground truth are compared.\\
{\bf Results:} The retrospective experiments include 181 clinical CMR exams, which are randomly split into training, validation, and test sets in the ratio of 64:16:20.
Our system yields the mean angular difference and point-to-plane distance of 5.68$^\circ$ and 3.12 mm, respectively, on the held-out test set.
%, in a retrospective study of 181 clinical CMR exams.
It not only achieves superior accuracy to existing approaches including conventional {\color{red}atlas-based} and newer {\color{red}deep-learning-based} in prescribing the four standard CMR planes {\color{red}but also} demonstrates prescription of the first {\color{red}cardiac-anatomy-oriented} plane(s) from the {\color{red}body-oriented} scout.\\
{\bf Conclusions:} The proposed system demonstrates accurate automatic CMR view plane prescription based on deep learning on properly archived data, without the need for further manual annotation.
This work opens a new direction for automatic view planning of anatomy-oriented medical imaging beyond CMR.\\
\end{abstract}
%\note{This is a sample note.}

%\newpage     %may or may not be needed

%The table of contents is for drafting and refereeing purposes only. Note
%that all links to references, tables and figures can be clicked on and
%returned to calling point using cmd[ on a Mac using Preview or some
%equivalent on PCs (see View - go to on whatever reader).
%\tableofcontents

\newpage

\setlength{\baselineskip}{0.7cm}      %double spacing		

\pagenumbering{arabic}
\setcounter{page}{1}
\pagestyle{fancy}
\section{Introduction}
\label{sec:intro}
Heart disease is the leading cause of death in the United States, accounting for approximately one in four deaths\cite{murphy2018mortality}.
The most common type of heart disease is {\color{red}ischaemic heart disease}, the leading cause of global disability-adjusted life years (DALYs) in age groups 50--74 years and 75+ years\cite{vos2020global}.
Cardiac magnetic resonance (CMR) imaging is the gold standard for the quantification of volumetry, function, and blood flow of the heart\cite{la2012cardiac}.
%both anatomical and functional imaging of the heart
%For example, cine cardiac MRI allows accurate measurement of the left ventricular ejection fraction (LVEF), which (describe what it is) and is frequently used in clinical practice.
A great deal of effort is put into the development of algorithms for accurate, robust, and automated analysis of CMR images, with a central focus on the segmentation of cardiac structures\cite{bai2019self,painchaud2019cardiac,robinson2017automatic,wei2013comprehensive,zotti2018convolutional,chen2019learning}.
%wei2015medical or registration of multiple sequences and/or time points [?,?,?].
However, much less attention has been paid to automatic view planning for the acquisition of CMR, {\color{red}which remains} {\color{cyan}a demanding task}
%challenging
in clinical practice.
First, imaging planes of CMR are customized for each individual based on specific cardiac structural landmarks\cite{kramer2020standardized},
%including the mitral valve, apex, \emph{etc}
and the planning process demands specialist expertise\cite{suinesiaputra2015quantification}.
Second, the planning process adopts a multi-step approach involving several localizers defined by the cardiac anatomy, which is complex, time-consuming, and subject to operator-induced variations\cite{lu2011automatic}.
These factors may constrain the use of CMR in clinical practice. %{\color{cyan}(e.g., the throughout of a deployed facility can be limited by the availability of qualified operators)}.
Hence, {\color{red}an} automatic {\color{red}planning} system is expected to increase the impact of the specific imaging technology on {\color{red}the} care of patients suffering from cardiovascular diseases.

\begin{figure}[t]
  \begin{center}
  \includegraphics[width=8cm]{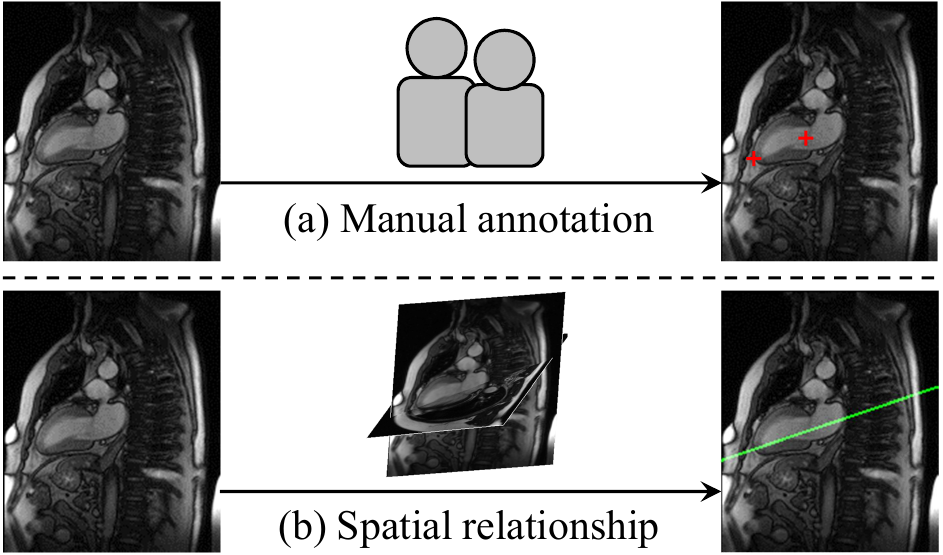}
  \captionv{12}{}{Conceptual comparison of our proposed approach with previous CNN-based methods in terms of training annotation.
  (a) Existing methods (e.g., Blansit et al.\cite{blansit2019deep}) relied on manual annotations of key landmarks, such as the apex and mitral valve (marked with
  %red
  crosses) in the vertical long-axis localizer, to train the deep CNNs.
  (b) Our approach, in contrast, mines the spatial relationship between intersecting views for supervision by the intersecting line, % (green)
  thus eliminating the need for the manual annotations.
  Note that both approaches (Blansit et al.'s and ours) no longer require manual annotations for view planning once the models are trained.\label{fig:concept}}
  \end{center}
\end{figure}

Few works attempted automatic view planning for CMR\cite{lu2011automatic,frick2011fully,alansary2018automatic,blansit2019deep}.
For example, both Lu et al.\cite{lu2011automatic} and Frick et al.\cite{frick2011fully} tackled this challenging task from the perspective of classical atlas-based methods by fitting triangular mesh-based heart models into a 3D volume.
%Lu \emph{eh al.} proposed to fit a triangular mesh model for the left ventricle (LV) and LV outflow tract (LVOT) to localize and delineate cardiac anatomies in a 3D volume, and detect a set of cardiac landmarks to anchor chambers for view prescription\cite{lu2011automatic}.
%Similarly, Frick et al. proposed surface model based segmentation and registration of atlas landmarks for fully automatic geometry planning for CMR\cite{frick2011fully}.
Later, Alansary et al.\cite{alansary2018automatic} proposed to employ reinforcement learning to prescribe the standard four-chamber long-axis CMR plane from a thoracic volume.
%where an agent moved a randomly initiated plane to gradually approach the ground truth plane.
A common foundation of these works was the use of a 3D magnetic resonance imaging (MRI) volume from which the standard CMR planes were prescribed.
%What these works had in common was that they assumed the availability of a 3D thoracic MRI volume from which the standard CMR views were prescribed.
However, such a 3D volume is not typically acquired in {\color{red}the} current clinic routine,
%However, this assumption is incompatible with the existing clinical work flow,
where the standard CMR planes are sequentially optimized based on a set of 2D localizers.
%involves a multi-step approach to sequentially optimize standard CMR view planes based on a series of double-oblique 2D localizers
%incrementally plans the next view based on a (series of) 2D view(s)/localizer(s) already planned%环环相扣
{\color{red}To} develop a clinic-compatible system, Blansit et al.\cite{blansit2019deep} proposed to sequentially prescribe standard CMR planes given a vertical long-axis localizer (also known as the pseudo two-chamber (p2C) localizer), driven by {\color{red}deep-learning-based} localization of key landmarks.
However,
%to exploit the power of deep convolutional neural networks (CNNs),
this method relied on extensive manual annotations of key anatomical landmarks (Fig. \ref{fig:concept}(a)) to train the deep convolutional neural networks (CNNs), though the manual annotations were no longer needed for plane {\color{red}prescriptions} after the model was trained.
Besides,
%as the authors pointed out,
the p2C localizer---the starting point of the entire system---was assumed given.
Yet in practice{\color{red},} it requires expertise in cardiac anatomy to prescribe this localizer from scout images in normal body planes (e.g., the axial view), which is an obstacle to a more automated pipeline.

In this work, we propose a clinic-compatible, automated system for the complete pipeline of CMR view planning.
{\color{red}Our system takes advantage of the power of deep CNNs---similar to Blansit et al.\cite{blansit2019deep}---but eliminates the need for annotation by mining data properties}
%via self-supervised learning
\cite{bai2019self,jing2020self}. %,zhou2019models
Above all, we make a critical observation that the way how the existing CMR data have been prescribed is self-contained in correctly recorded data, e.g., in the Digital Imaging and Communications in Medicine (DICOM) format, in the form of the spatial relationship between views.
Then, inspired by the recent progress in keypoint-based object detection\cite{duan2019centernet,law2018cornernet,zhou2019bottom}, %zhou2019objects,
we propose to regress the intersecting lines between the views, which can be readily computed using the spatial relationship (Fig. \ref{fig:concept}(b)).
%The extension from the detection of single point to that of a line is natural for our context here, as the intersection lines are not only a collection of points but more importantly defined by key cardiac landmarks.
Training the networks to predict these intersecting lines {\color{red}teaches} them to reason about the key cardiac landmarks that defined these lines when the operators produced the CMR data, while at the same time {\color{red}eliminating} the need for manual annotation of the landmarks.
Our secondary observation is that multiple target planes that are prescribed in the same source view often have fixed (though rough) spatial {\color{red}relationships concerning} each other, e.g., the standard short- and long-axis planes should appear close to perpendicular in a long-axis localizer, instead of overlapping or being parallel.
Hence, we propose to utilize the interplay of the multiple target planes predicted in a source view via a stacked hourglass network\cite{newell2016stacked} featuring repeated top-down and bottom-up processing, to gradually improve the predictions through the hourglasses.
%Our work is motivated by the recent surge of self-supervised learning [?,?,?] and built upon the recent progress in CNN-based key point detection [?,?,?].
%Second, we take a step forward from existing keypoint-detection networks, by extending the detection of a single point to that of the intersecting line (\emph{i.e.}, a collection of points) between different image view planes.
%The process of the network predicting the intersecting line of a CMR view within a given localizer essentially mimics the behavior of an operator when planning a specific view based on the cardiac landmarks in a localizer.
After the intersecting lines are predicted in the localizers, the standard CMR planes are eventually planned by aggregating the predictions in multiple localizers to search for a globally optimal prescription.

In summary, our contributions are as follows:
\begin{itemize}
\item First, we propose a CNN-based, {\color{red}clinical-compatible} system for automatic CMR view planning, which eliminates the need for manual annotation via {\color{yellow}supervision by}
    %self-supervised learning of
    the spatial relationship between views.
\item Second, we propose to employ the stacked hourglass architecture to make use of the spatial interplay of the multiple target planes that are predicted in the same source view, for better regression results.
\item Third, we propose to aggregate multi-source-view information for globally optimal plane prescription, to mimic the clinical practice of multi-view planning.
\item Fourth, we conduct extensive experiments to study the proposed system and demonstrate its competence/{\color{red}superiority} to existing methods.
\item Last but not least, we demonstrate the prescription of the {\color{red}cardiac-anatomy-defined} localizers (including the p2C) given the axial localizer, bridging the gap between general {\color{red}body-oriented} and customized {\color{red}cardiac-anatomy-oriented} planes for a more automated pipeline.
\end{itemize}

{\color{purple}This work substantially expands our preliminary exploration reported in a conference proceedings paper\cite{wei2021training} in two main aspects: (i)~it exploits the interplay of multiple target planes predicted in a source view via the stacked hourglass architecture, to effectively reduce the mean normal deviation and point-to-plane distance by $\sim$5\% and $\sim$10\%, respectively, and (ii) it extends and demonstrates the application of the proposed approach to diverse imaging protocols (i.e., parallel- and single-view planning) for broad clinical applicability.}
%The rest of this paper is organized as following.
%Section \ref{sec:preliminary} introduces the relevant terminology of CMR, and the imaging protocol of the CMR data used in this study, as a preliminary.
%Section \ref{sec:method} describes our proposed approach to automatic CMR view planning in details.
%Section \ref{sec:experiment} conducts thorough experiments on a set of real clinical data, followed by the discussion and conclusion in Section \ref{sec:conclusion}.

\begin{figure}[t]
  \begin{center}
  \includegraphics[width=13cm]{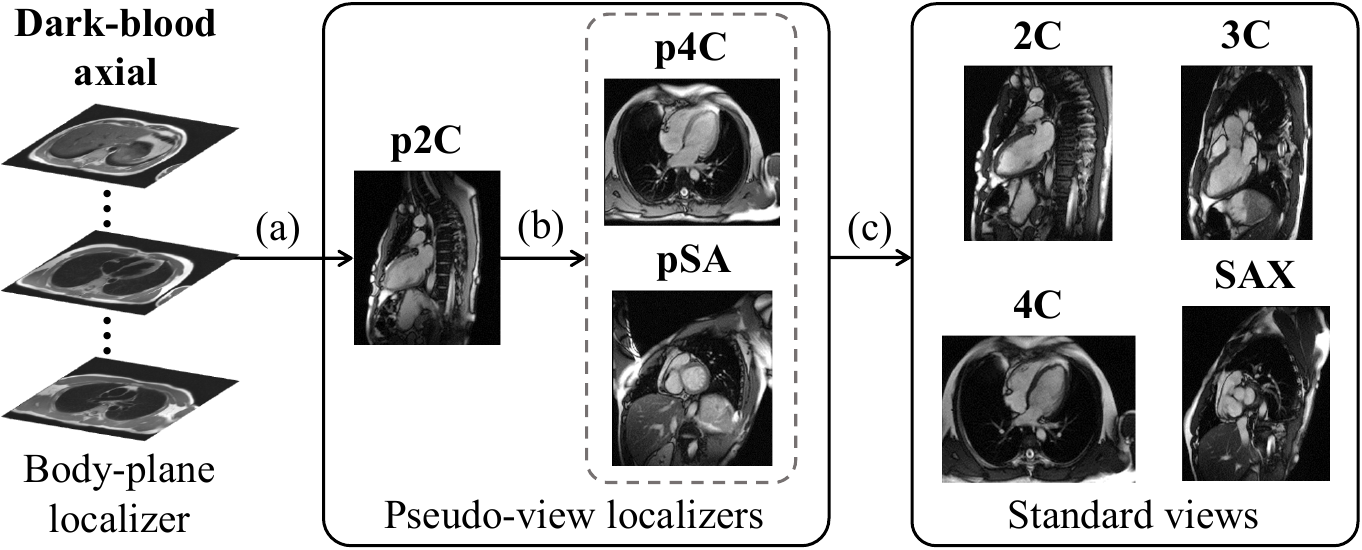}
  \captionv{16.5}{}{\color{cyan}Our imaging protocol given the body-plane oriented (axial), multi-slice dark-blood localizer.
  (a) A single-slice {\color{red}pseudo-2C} (p2C) localizer---the first {\color{red}cardiac-anatomy-oriented} view---is prescribed from the axial localizer.
  (b) A single-slice {\color{red}pseudo-4C} localizer (p4C) and a multi-slice {\color{red}pseudo-SAX} (pSA) localizer are prescribed based on the p2C localizer.
  (c) Standard 2C, 3C, 4C, and SAX views are optimized from the set of pseudo-view localizers.\label{fig:protocol}}
  \end{center}
\end{figure}

\section{Preliminaries}\label{sec:preliminary}

\subsection{Imaging Protocol and Problem Statement}\label{sec:preliminary:protocol}
Different from the commonly used axial, sagittal, or coronal plane oriented with respect to the long axis of the body (the body planes), CMR adopts a set of double-oblique, cardiac anatomy defined imaging planes customized for each individual.
Often these planes are prescribed along the long axis (LAX) or short axis (SAX) of the left ventricle (LV) and {\color{red}concerning} specific cardiac structural landmarks
(e.g., the apex and mitral valve),
%(the apex, mitral valve, aortic valve, \emph{etc.}),
for optimal visualization of structures of interest and evaluation of cardiac functions.
The {\color{red}most} used standard CMR planes {\color{cyan}(Fig. \ref{fig:protocol} rightmost)} include the LV two-chamber (2C) LAX, three-chamber (3C) LAX, four-chamber (4C) LAX (also known as the vertical, LV outflow tract (LVOT), and horizontal LAX planes, respectively), and SAX planes, which provide complementary information for a comprehensive evaluation of the heart.
%(i) the LV two chamber (2C) view for evaluating the anterior and inferior walls and apex of the left ventricle,
%(ii) the three chamber (3C) view showing the aortic root and aortic valve, left ventricular outflow tract, mitral valve, and the anteroseptal and inferolateral walls of the left ventricle
% the four chamber (4C) view for
% the short axis (SAX) view ,

Our imaging protocol generally follows Kramer et al.\cite{kramer2020standardized} for LV structure and function.
After adjusting the heart to the isocenter of the bore, a multi-slice axial dark-blood localizer is firstly prescribed through the chest from sagittal and coronal scouts, serving as the basis for the following {\color{red}cardiac-anatomy-oriented} views {\color{cyan}(Fig. \ref{fig:protocol} leftmost)}.
Next, a single-slice {\color{red}pseudo-2C} (p2C) localizer---the first {\color{red}cardiac-anatomy-oriented} view---is prescribed from the axial localizer {\color{cyan}(Fig. \ref{fig:protocol}(a))}.
Then, a single-slice {\color{red}pseudo-4C} localizer (p4C) and a multi-slice {\color{red}pseudo-SAX} (pSA) localizer are prescribed based on the p2C localizer {\color{cyan}(Fig. \ref{fig:protocol}(b))}.
Although similarly defined by the cardiac anatomy, these localizers usually cannot provide accurate anatomic and functional characterization of the heart like the standard-plane views due to the obliquity of the heart walls to the body planes\cite{ginat2011cardiac}, hence are often referred to as the ``pseudo'' LAX (pLA) and SAX views\cite{10.1093/eurheartj/ehw680}.
Eventually, the p2C, p4C, and pSA localizers are used to sequentially optimize the imaging planes for the standard 2C, 3C, 4C, and SAX views  {\color{cyan}(Fig. \ref{fig:protocol}(c))}.
%Eventually, the standard 2C, 3C, 4C, and SAX views are prescribed from the VLAX, HLAX, and SAX localizers.
%Note that the imaging planes of the standard LAX and SAX views are different from those of the localizers with the same names.
%Rather, the latter were used to
%localize the long- and short-axis of the heart and
%plan the former.
For clarity, we refer to an imaging plane to prescribe as the \emph{target} plane, and the view(s) leading to the specific target plane as its \emph{source} view(s) {\color{red}afterward}.
%{\color{yellow}The complete target-and-source dependencies for our imaging protocol are summarized in Table \ref{tab:view_dependency};
%besides, Figs. \ref{fig:loc} and \ref{fig:framework} illustrate the dependencies for the localizers and the standard LAX and SAX views, respectively.}

{\color{cyan}In this work, besides planning the standard LAX and SAX planes (Fig. \ref{fig:protocol}(c); only the most basal SAX plane is considered) as in the literature\cite{lu2011automatic,alansary2018automatic,blansit2019deep,frick2011fully}, we also investigate the prescription of the pseudo-view localizers from the axial localizer (Figs. \ref{fig:protocol}(a) and (b)).
Especially, as the first step of deriving {\color{red}cardiac-anatomy-oriented} imaging planes from the {\color{red}body-oriented} planes, prescribing the p2C localizer (Fig. \ref{fig:protocol}(a))
%requires expertise for identification of specific cardiac structural landmarks and
is indispensable for a clinic-compatible system yet
%Yet, this step has not been
remains uninvestigated in existing literature\cite{blansit2019deep}.}

\subsection{Basic Geometry}
{\color{cyan}DICOM\footnote{https://www.dicomstandard.org/} is the international standard for medical images and related information, and is used by almost all radiographs nowadays.
It defines the formats for medical images that can be exchanged with the data and quality necessary for clinical use.
As shown in Fig. \ref{fig:geometry}(a), the DICOM header contains two attributes that record the location and orientation of a medical image:
(i) ImagePositionPatient (IPP): the $x$, $y$, and $z$ coordinates of the top left
corner (center of the first pixel transmitted) of the image with respect to the \emph{reference coordinate system} (RCS), a patient-based 3D coordinate system;
and (ii) ImageOrientationPatient (IOP): the direction cosines of the first row and the first column of the image with respect to the RCS.
Using IPP and IOP, the spatial relationship between any two radiographs (of the same
scan session) can be readily derived, including their intersecting line, if existing.\footnote{Due to space limit, we do not include the detailed derivation in this paper.}

\begin{figure}[t]
  \begin{center}
  \includegraphics[width=13cm]{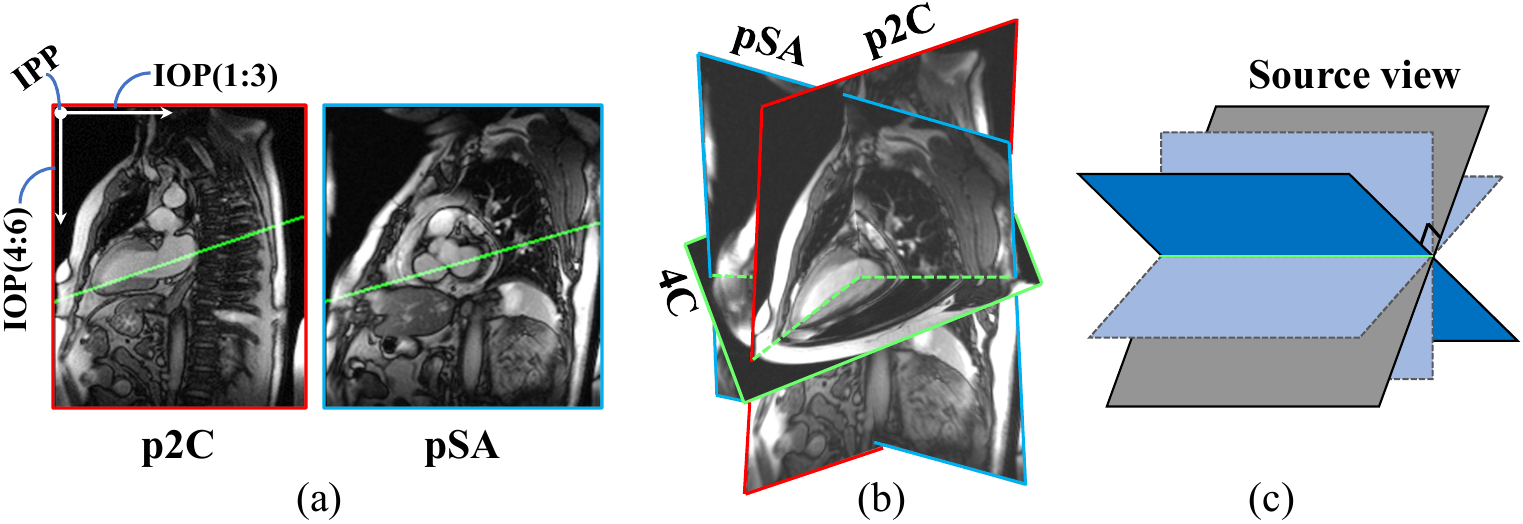}
  \captionv{16.5}{}{\color{cyan}Geometry for target plane prescription. (a) Prescription of the 4C plane given the p2C and pSA localizers
  %(outlined in red and cyan, respectively),
  (represented by its intersecting lines
  %(in green)
  with the source view localizers).
  The target plane is prescribed according to the individual cardiac anatomy: the 4C plane is prescribed from the p2C localizer through the apex and center of the mitral and tricuspid valves.
  This can be modified and/or cross-checked on basal pSA localizer, to have the plane cross the acute margin of the right ventricular (RV) free wall and perpendicular to the interventricular septum\cite{kramer2020standardized}.
  (b) 3D visualization of the spatial relationship shown in (a).
  (c) Single-view prescription illustrated with a toy plot:
  %(gray for the source view and dark blue for the target plane perpendicular to the source).
  the target plane (in solid outline) is presumed to be orthogonal to the source view. \label{fig:geometry}}
  \end{center}
\end{figure}

Besides the RCS, another involved coordinate system is the 2D \emph{image coordinate system} (ICS), with the top left corner pixel of the image as the origin, and the first row and column as the $x$ and $y$ axes, respectively.
A line in the ICS can be denoted by $Ax+By+C=0$, where $(x,y)$ denote pixel coordinates and $(A, B, C)$ are linear parameters.

\subsection{Single- and Multi-View Target Plane Prescription}\label{sec:preliminary:multiview}
In clinical practice, given one or more source views, technicians prescribe a target plane in accordance with the specific, individual cardiac anatomy.
Take the prescription of the standard 4C plane given the p2C and pSA localizers for example (Fig. \ref{fig:geometry}(a)--(b)):
the 4C plane is prescribed from the p2C localizer through the apex and center of the mitral and tricuspid valves.
This can be modified and/or cross-checked on basal pSA localizer, to have the plane cross the acute margin of the right ventricular (RV) free wall and perpendicular to the interventricular septum\cite{kramer2020standardized}.
%``The 4-chamber long-axis view is prescribed from the 2-chamber long-axis view through the apex and center of the mitral and tricuspid valves.
%This can be modified and/or cross-checked on basal short-axis views, to have the plane cross the acute margin of the right ventricular (RV) free wall and perpendicular to the interventricular septum.''\cite{kramer2020standardized}
Therefore, the target plane is prescribed at an overall preferable location observed in multiple source views.
It is worth mentioning that in multi-view prescription, the multiple source views can be either intersecting (e.g., the p2C and pSA localizers in Fig. \ref{fig:geometry}(a)--(b)) or parallel (e.g., the multi-slice axial localizer in Fig. \ref{fig:protocol} left) with each other, or sometimes both.

%\begin{figure}[t]
%  \centering\footnotesize
%  \includegraphics[width=.95\columnwidth]{geometry}\\
%  \mbox{(a) p2C\hspace{1.5cm}(b) pSA\hspace{2.3cm}(c)\hspace*{1.5cm}}
%  \caption{\color{cyan}(a)--(b): Prescription of the 4C plane given the p2C and pSA views, represented by its intersecting lines (in red) with the source views.
%  As we can see, the target plane is prescribed in accordance with the individual cardiac anatomy: ``The 4-chamber long-axis view is prescribed from the 2-chamber long-axis view through the apex and center of the mitral and tricuspid valves.
%  This can be modified and/or cross-checked on basal short-axis views, to have the plane cross the acute margin of the right ventricular (RV) free wall and perpendicular to the interventricular septum.''\cite{kramer2020standardized}}\label{fig:geometry}
%\end{figure}

As to the geometry of single-view prescription, we use the toy plot in Fig. \ref{fig:geometry}(c) for explanation without loss of generality.
As we can see, prescribing the target plane within a single source view
%(shaded in gray)
is indecisive to determine a unique plane, as there exist indefinite candidate planes (in dashed outlines)
%(shaded in light blue)
rotating about the only prescribing line. %(shown in green).
In this case, the target plane (in solid outline)
%(shaded in dark blue)
is often presumed to be orthogonal to the source view, to resolve the ambiguity in its 3D orientation.
%Alternatively, additional source views can be incorporated to resolve the ambiguity, where the target plane is prescribed at (overall) preferable locations in multiple source views.
%In this latter case, the multiple source views can be either parallel (such as the multi-slice axial localizer in Fig. \ref{fig:protocol}) or intersecting (such as Fig. \ref{fig:geometry}(a)--(b)) with each other, or sometimes both.
}

%\subsection{SSL and multi-view aggregation based CMR view planning}
\section{Methods}\label{sec:method}
\begin{figure}[tb]
\begin{center}
    \includegraphics[width=\textwidth,,trim=0 0 0 0,clip]{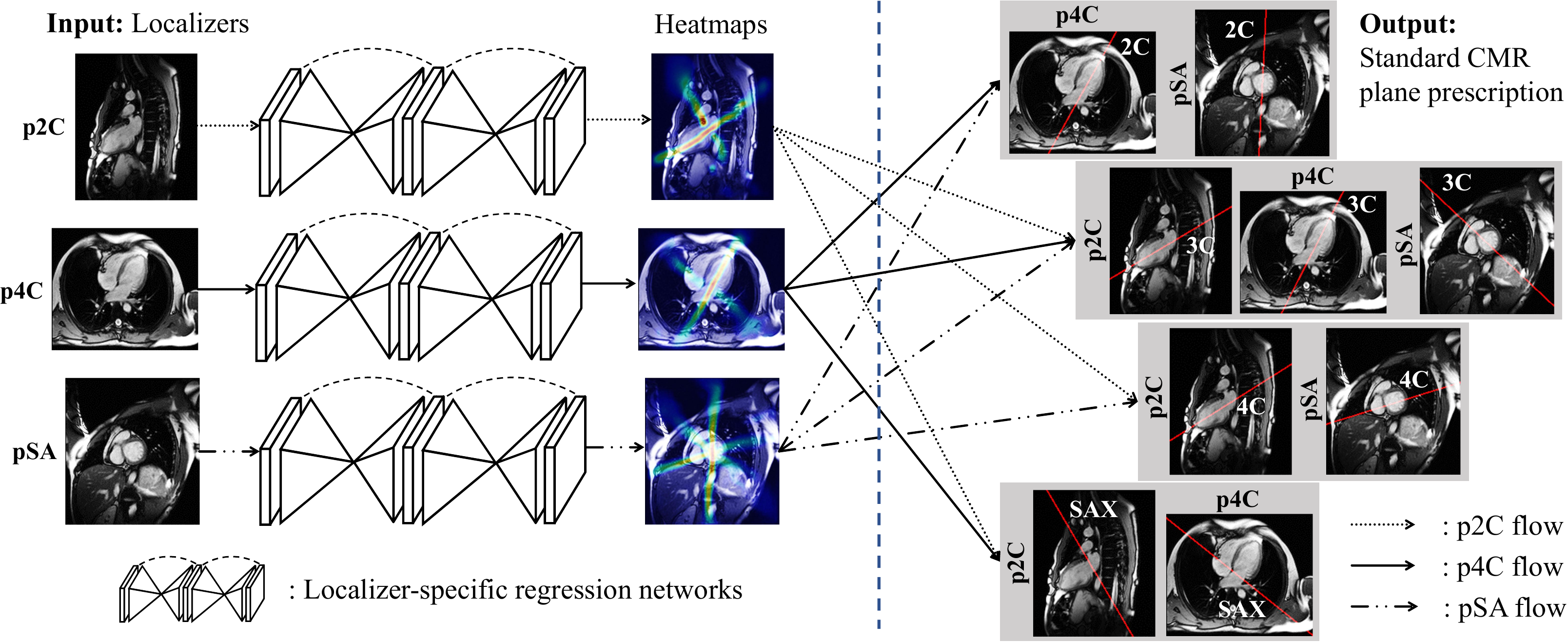}
    \captionv{16.5}{}{Overview of the proposed approach to automatic CMR view planning (illustrated with the task of planning the standard CMR planes from the localizers).
    {\color{cyan}The predictive information flows of the localizers are differentiated with different line styles (legend on the bottom right).}
    Left: prediction of standard planes in localizers via heatmap regression with stacked hourglass networks.
    {\color{cyan}A separate regression model is trained for each localizer.}
    Right: prescription of standard planes by aggregating heatmaps predicted in multiple localizers.
    {\color{cyan}\emph{Note that the prescribed planes are represented by prescription lines
    %(red)
    in the input localizers.}
    As we can see, the prescription lines pass meaningful anatomical landmarks just like manual prescriptions do, e.g., the standard 2C plane passes the apex and center of the mitral valve in the p4C localizer.}\label{fig:framework}}
\end{center}
\end{figure}

The overview of our proposed system is shown in Fig. \ref{fig:framework}, including two main steps.
First, given a set of localizers as input, heatmaps for the target planes are predicted by the view-specific regression networks.
Second, multiple heatmaps for a specific target plane are aggregated to prescribe the target plane as output.
Below we elaborate {\color{red}on} the details.

\subsection{{\color{red}Cardiac-Anatomy-Aware} Target Plane Regression}% with Self-Supervised Learning
A previous work on CNN-based automatic CMR view planning regressed locations of the cardiac structural landmarks {\color{red}and} prescribed the standard planes by connecting the landmarks\cite{blansit2019deep}.
A drawback was that the landmark regression networks needed extensive annotations for training.
In this work, we propose to mine the spatial relationship among the CMR data {\color{red}and} directly regress the intersecting lines of the target plane with the source views.
As the spatial information of each CMR slice is recorded in its data header (e.g., DICOM tags), we can readily compute the intersecting lines.
Therefore, the supervising ground truth of our proposed regression task is self-contained in the data, requiring no additional manual annotation.
In addition, the intersecting lines are actually the prescription lines prescribed by the technologist at the time of image acquisition using cardiac landmarks, and retrospectively identified from the spatial relationship.
Hence, our networks are still trained to learn the cardiac anatomy by the proposed regression task, while supervised by technologist prescribed planes.

For practical implementation, we train the networks to regress a heatmap
%(see Fig. \ref{fig:framework} left for examples of regressed heatmaps)
defined by the distance to the intersecting line, instead of the line itself.
This strategy is commonly adopted in the keypoint detection literature
\cite{duan2019centernet,law2018cornernet,zhou2019bottom,pfister2015flowing}, %zhou2019objects,
where the regression target is the heatmaps defined by the keypoints.
The benefits of regressing heatmaps include better interpretability and ease of learning for the network\cite{pfister2015flowing}.
{\color{cyan}Specifically, using the ``softened'' ground truth as the training target  imposes  less  penalty  when  the prediction gets closer to the exact location, thus encouraging the prediction to gradually approach the desired status.}
Formally, denoting the intersecting line in the \emph{2D image coordinate system of a source view} by $Ax+By+C=0$, where $(x,y)$ are coordinates and $(A, B, C)$ are parameters, the heatmap is computed as
\begin{linenomath*}
\begin{equation}\label{eq:heatmap}
    H(x,y)=\exp\big[-{(Ax+By+C)^2}\big/\big(2\sigma^2(A^2+B^2)\big)\big],
\end{equation}
\end{linenomath*}
where $\sigma$ is a hyperparameter {\color{cyan}controlling the extent of ``softening''}.
%denoting the Gaussian kernel size.
We define $\sigma$ {\color{red}concerning} the slice thickness of the target view for concrete physical meaning {\color{red}and} study its impact later with experiments.
{\color{cyan}Intuitively, the further a pixel is from the intersecting line, the cooler it is in the heatmap.}
An L2 loss is employed to train the network:
\begin{linenomath*}
\begin{equation}\label{eq:L_heat}
    \mathcal{L} = \frac{1}{T}\frac{1}{|\Omega|}
    {\sum}_{t=1}^T{\sum}_{(x,y)\in\Omega}
    \big\|H_t(x,y) - \hat{H}_t(x,y)\big\|^2,
\end{equation}
\end{linenomath*}
where $T$ is the total number of target planes prescribed from a source view (also the number of channels of the network's output), $(x,y)$ iterates over all pixels in the source view image domain $\Omega$, and $\hat{H}$ is the heatmap predicted by the network.
%(see Fig. \ref{fig:framework} middle for example predictions by trained networks)

{\color{cyan}As shown in Fig. \ref{fig:framework} left, we train a separate regression model for each localizer.
Also, the model is trained to regress the heatmaps corresponding to all the standard planes that are prescribed in a specific localizer in our clinical protocol, i.e., 3C, 4C, and SAX in p2C; 2C, 3C, and SAX in p4C; and 2C, 3C, and 4C in pSA.
Therefore, $T=3$ (Eqn. (\ref{eq:L_heat})) for the p2C, p4C, and pSA localizers.
Fig. \ref{fig:framework} middle shows example {\color{red}predictions} by trained networks,
%for each of the p2C, p4C, and pSA localizers,
where the heatmaps for different target planes are overlaid.\footnote{\color{cyan}Due to the closeness (sometimes coincidence) of the standard 3C plane with the standard 4C (2C) plane in the p2C (p4C) localizer, only two stripes of high response can be observed in the overlaid heatmaps predicted in the p2C (p4C) localizer, instead of three as in the pSA localizer.}
}

%Note that it is generally infeasible to reliably predict the intersecting line between the same group of localizer and cine planes (e.g., between the 2C localizer and 2C cine view), as the intersection line is not anatomically meaningful.

\subsection{Stacked Hourglass Networks with Deep Supervision}
As shown in Fig. \ref{fig:framework}, often multiple standard CMR planes are regressed in each localizer.
The interplay of the planes regressed in the same source view
%---\emph{i.e.}, concordance and contradiction---
can provide valuable clues about the structured regression tasks.
For example, the 4C and SAX planes should intersect {\color{red}at} a large angle in the p2C localizer, instead of being parallel or even overlapping considering anatomic impossibility.
To make use of the clues, we employ the stacked hourglass network (SHG-Net)\cite{newell2016stacked} as our regression network.
SHG-Net features repeated bottom-up, top-down processing in conjunction with intermediate supervision, and was originally proposed for the task of human pose estimation exploiting the relations among predicted joint locations.
Concretely, a single hourglass is a U-Net-style encoder-decoder structure {\color{red}and} produces a set of intermediate heatmaps for loss computation.
Then, the intermediate heatmaps are remapped back to the feature space with 1$\times$1 convolutions, and fused with the intermediate features output by preceding hourglasses (including itself),
%the hourglass itself and its predecessor
%its input (\emph{i.e.}, the features output by the preceding hourglass)
via element-wise addition.
Next, the fusion result is used as the input to the next hourglass, and so forth.
For more details, readers are referred to Newell et al.\cite{newell2016stacked}
In this way, the initial estimates and high-level features are reevaluated with respect to each other, leading to iterative refinement of the regressed planes.
%`The precise position of a joint is an indispensable cue for other decisions being made by the network. With a structured problem like pose estimation, the output is an interplay of many different features that should come together to form a coherent understanding of the scene. Contradicting evidence and anatomic impossiblity are big giveaways...'

As the predicted heatmaps are expected to gradually improve from earlier to later hourglasses, we accordingly set larger weights for the predictions of later hourglasses and define the total loss to optimize as
\begin{linenomath*}
\begin{equation}\label{eq:L_hourglass}
  \mathcal{L} = {\sum}_{n=1}^N w_n\mathcal{L}_n \Big/ {\sum}_{n=1}^N w_n,
\end{equation}
\end{linenomath*}
where $N$ is the number of hourglasses, $\mathcal{L}_n$ is the L2 loss (Eqn. (\ref{eq:L_heat})) on the heatmaps predicted by the $n$\textsuperscript{th} hourglass, and $w_n=1\big/2^{N-n}$.
Note that the intermediate heatmaps are only used for the purpose of training supervision, and only the heatmaps predicted by the last hourglass are used for the CMR plane prescription described next.

\subsection{Plane Prescription with Multi-View Aggregation}
After the heatmaps are predicted for all the source views of a specific target plane, we prescribe the target plane by aggregating information from all these heatmaps.
Specifically, the plane with the greatest collective response aggregated over its intersecting lines with all the source heatmaps is prescribed.
%we mimic the clinical practice where skilled operators plan the CMR views by checking the locations of a candidate plane within multiple source views.
Formally, let us denote a candidate plane in the RCS by $P=(p, \theta, \phi)$, where $p$ is a point on the plane, $\theta$ and $\phi$ are the polar and azimuthal angles of its normal in the spherical coordinate system;
and denote the intersecting line segment $l_v$ between $P$ and the $v$\textsuperscript{th} source view (of $V$ total source views) by the exhaustive set of pixels lying on the line within the source view image $I_v$: $l_v=\{(x, y)|(x, y)\in P\cap I_v\}$, where $P\cap I_v$ denotes the intersection.
{\color{cyan}Then, the optimal target plane can be formally defined as
\begin{linenomath*}
\begin{equation}\label{eq:grid_search}
 \hat{P}=\operatorname{argmax}_{(p,\theta,\phi)}S,\ \text{where }
 S={\sum}_{v=1}^{V}{\sum}_{(x,y)\in l_v}\hat{H}_v(x,y).
\end{equation}
\end{linenomath*}
%As the predicted heatmaps are quite reasonable, we only need to
We propose to grid search for the optimal triplet $(\hat{p},\hat{\theta},\hat{\phi})$ with two strategies to reduce the search space (Algorithm \ref{alg:grid_search}).
First, we constrain the search for $\hat{p}$ along a line segment $l_s$ with which the target plane should intersect.}
In practice, we use the intersecting line segment between two intersecting source views as $l_s$.
%\footnote{\color{yellow}if available, or simply any row and column of the source view image(s), can be used.}
Second, a three-level coarse-to-fine pyramidal search is employed, with the steps set to 15, 5, and 1 pixel(s) for $p$,  and 15, 5, and 1 degree(s) for $\theta$ and $\phi$, respectively.
Note that our multi-view aggregation naturally takes into account the networks' confidence in the predictions, where higher values in the regressed heatmaps indicate higher confidence.

{\color{cyan}According to our clinical protocol, we prescribe the 2C plane from the p4C and pSA localizers; the 3C plane from the p2C, p3C, and p4C localizers; the 4C plane from the p2C and pSA localizers; and the SAX plane from the p2C and p4C localizers (Fig. \ref{fig:framework} right), using the multi-view aggregation.}

\begin{algorithm}[t]\color{cyan}
\SetKwInOut{Init}{Init}
\KwIn{the set of predicted heatmaps $\{\hat{H}_v\}$ for all source views, line segment $l_s$ along which $\hat{p}$ is searched for}
\Init{set $p_\mathrm{low}$ and $p_\mathrm{high}$ to the first and last points on $l_s$, $\theta_\mathrm{low}=\phi_\mathrm{low}=0^\circ, \theta_\mathrm{high}=\phi_\mathrm{high}=179^\circ$; $\hat{S}=0$}
\For{$s$ in $[15, 5, 1]$}{
    \For{$p\leftarrow p_\mathrm{low}$ \KwTo $p_\mathrm{high}$ with step $s$}{
        \For{$\theta\leftarrow \theta_\mathrm{low}$ \KwTo $\theta_\mathrm{high}$ with step $s$}{
            \For{$\phi\leftarrow \phi_\mathrm{low}$ \KwTo $\phi_\mathrm{high}$ with step $s$}{
                find the intersecting line segment $l_v$ between $P=(p,\theta, \phi)$ and every source view\;
                compute $S$ in Eqn. (\ref{eq:grid_search})\;
                \If{$S>\hat{S}$}{
                    \tcp{iteratively update max aggregation score $\hat{S}$ and record optimal $(\hat{p}, \hat{\theta}, \hat{\phi})$}
                    $\hat{S}\leftarrow S$, $(\hat{p}, \hat{\theta}, \hat{\phi})\leftarrow (p, \theta, \phi)$\;
                }
            }
        }
    }
    \tcp{update search range for the next pyramidal level, centered around optimal values of the current level}
    $(p_\mathrm{low}, p_\mathrm{high})\leftarrow(\hat{p}-s, \hat{p}+s)$, $(\theta_\mathrm{low}, \theta_\mathrm{high})\leftarrow(\hat{\theta}-s, \hat{\theta}+s)$, $(\phi_\mathrm{low}, \phi_\mathrm{high})\leftarrow(\hat{\phi}-s, \hat{\phi}+s)$ subject to boundary condition\;
}
\KwOut{optimal plane $\hat{P}=(\hat{p}, \hat{\theta}, \hat{\phi})$}
\caption{\sf Plane prescription via multi-view aggregation\label{alg:grid_search}}
\end{algorithm}

\subsection{Applicability to Diverse Clinical Protocols}
\begin{figure}[t]
\begin{center}
    \includegraphics[width=\textwidth,,trim=0 0 0 0,clip]{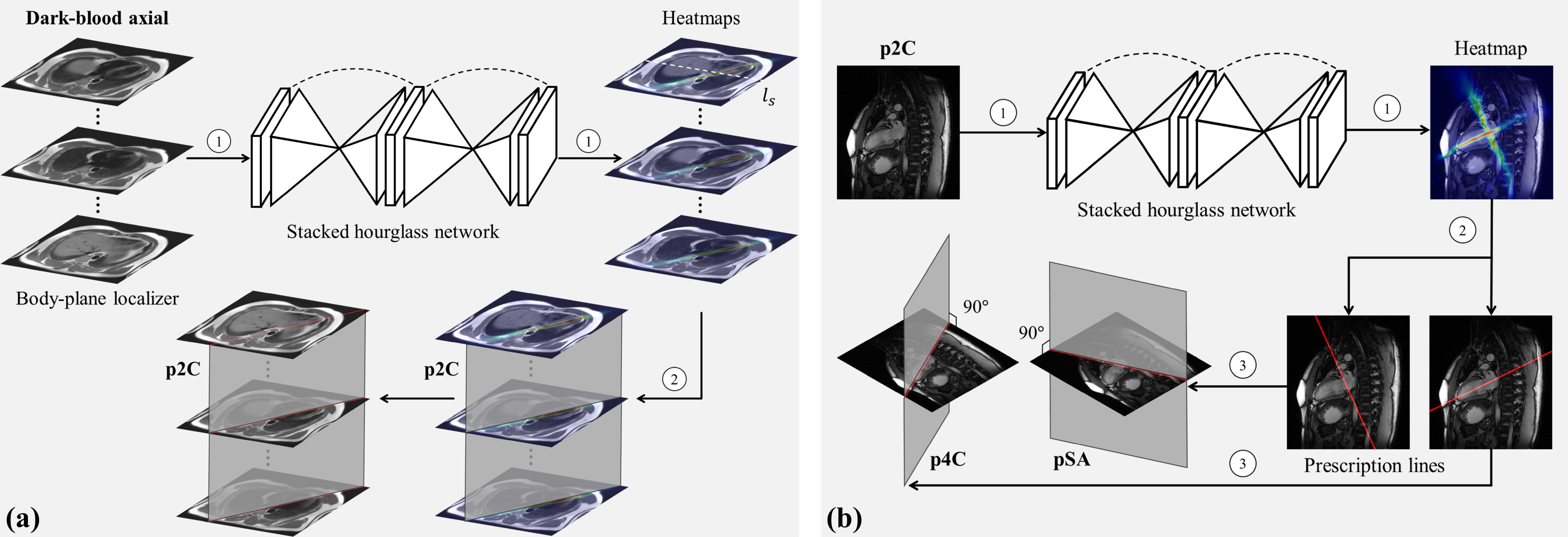}
    \captionv{16.5}{}{(a) Prescribe the p2C localizer plane from the multi-slice dark-blood axial localizer (i.e., step (a) in Fig. \ref{fig:protocol}): \textcircled{\small1} predict the heatmaps for all slices using the stacked hourglass network (SHG-Net), and \textcircled{\small2} grid search for the optimal plane (shaded in semi-transparent gray) whose collective response aggregated over its intersecting lines with all the heatmaps is the greatest (Algorithm 1).
    (b) Prescribe the p4C and pSA localizer planes from the p2C (i.e., step (b) in Fig. \ref{fig:protocol}): \textcircled{\small1} predict the p2C heatmap with the SHG-Net, \textcircled{\small2} grid search for two optimal prescription lines (one for p4C and one for pSA) whose collective responses aggregated along the line segments in the heatmap are the greatest (Eqn. (\ref{eq:line_search})), and \textcircled{\small3} prescribe the localizer planes (shaded in semi-transparent gray) passing through the optimal prescription lines while being orthogonal to the p2C plane.\label{fig:prescribe_locs}}
\end{center}
\end{figure}

The multi-view aggregation scheme presented above can be readily applied to a wide range of clinical protocols of CMR imaging, where the target plane is prescribed based on a set of source views.
For example, in the first step of our imaging protocol (Fig. \ref{fig:protocol}(a)), the p2C localizer is prescribed from the multi-slice axial localizer.
The axial localizer is different from the pseudo-view ones in three important ways: 1) it is a dark-blood sequence rather than bright-blood, 2) the multiple views therein are parallel (in contrast to the p2C, p4C, and pSA localizers which intersect with each other), and 3) it is body-oriented instead of cardiac anatomy oriented.
{\color{cyan}Yet the proposed aggregation scheme can be readily applied with a minor adaption: as the source views here do not intersect, we use the middle row of the first slice of the axial stack as $l_s$, to substitute for the absent intersecting line.}
Fig. \ref{fig:prescribe_locs}(a) illustrates the process of prescribing the p2C plane from the axial localizer.

Meanwhile, there also exist clinical protocols where a target plane is prescribed from a single source view.
{\color{red}For example, Blansit et al.\cite{blansit2019deep} prescribed the SAX stack from a single vertical LAX view}, {\color{yellow}and step (b) of our imaging protocol in Fig. \ref{fig:protocol} prescribes the p4C and pSA localizers from a single p2C view.}
In these cases, the target plane is {\color{yellow}often presumed to be orthogonal to the source view}, to resolve the ambiguity in its 3D orientation.
Then, the optimal plane problem in Eqn. (\ref{eq:grid_search}) degenerates into an optimal line problem, {\color{cyan}i.e., we only need to find a proper prescription line in the only source view.
Denoting the candidate prescription line in the image coordinate system of the source view by $L=(p, \theta)$}, where $p$ is a point on the line and $\theta$ is the line's planar directional angle, and its segment confined within the source view image $I$ by $l=\{(x,y)|(x,y)\in L\cap I\}$, we formally define the problem as
\begin{linenomath*}
\begin{equation}\label{eq:line_search}
  \hat{L}=\operatorname{argmax}_{(p,\theta)}S,\ \text{where }
 S={\sum}_{(x,y)\in l}\hat{H}(x,y).
\end{equation}
\end{linenomath*}
Again, the optimal pair $(\hat{p}, \hat{\theta})$ is grid searched {\color{cyan}to minimize $S$ in a three-level coarse-to-fine pyramid with the steps set to 15, 5, and 1 pixels (degrees), similar to Algorithm \ref{alg:grid_search}.}
Fig. \ref{fig:prescribe_locs}(b) illustrates the process of prescribing the {\color{red}p4C} and pSA planes from the p2C localizer.

In the experiments, we will demonstrate target plane prescription from both multiple parallel source views and a single source view with our proposed system {\color{cyan}(Section \ref{sec:exp:locs})}, to validate its applicability to diverse clinical protocols.

\section{Experiments and Results}\label{sec:experiment}
We conduct thorough experiments to evaluate the proposed {\color{red}clinically} compatible deep-learning system for automatic CMR view planning.
Concretely, the dataset and evaluation metrics used are first described, followed by implementation details of our system.
Then, important design considerations, including the hyperparameter $\sigma$,
%Gaussian kernel size,
multi-view aggregation, and number of stacked hourglasses, are empirically studied with the validation data.
Lastly, we evaluate the proposed system on the test data and compare its performance with that of existing methods.

\subsection{Dataset and Evaluation Metrics}
We retrospectively collected 181 CMR studies from 99 infarction patients (82 males and 7 females; age: 34--81 years;), {\color{yellow}with the approval of {\color{red}institution's} internal review board.}
{\color{purple}Of these patients, 82 had two exams and the rest had one.}
%92 (92.93\%) were male and 7 (7.07\%) were female (age range 34--81 years).
%---one at the time of infarction and the other three months later, whereas the rest had only one at infarction.
%The scans were performed on a 1.5T MRI system;
Details about the image acquisition parameters are provided in Table \ref{tab:protocol}.
%It is worth noting that the experimental data used in this work consist of all infraction cases of varying severity, potentially posing a big challenge to the proposed method.
The dataset is randomly split into training, validation, and test sets in the ratio of 64:16:20;
both exams (if available) of the same patient are assigned to the same set.
The validation set is used for empirical model optimization such as hyperparameter tuning.
%including hyperparameter tuning and model selection.
%determine hyper parameters, including number of training epochs and value of $\sigma$.
Then, the training and validation sets are combined to train a final model with optimized settings for evaluation on the held-out test set.
%optimized model is evaluated on the held-out test set.
The retrospectively identified planes prescribed by the technologists at the time of image acquisition are used as the ground truth.
Following existing literature\cite{lu2011automatic,alansary2018automatic,blansit2019deep,frick2011fully}, we use the angular deviation and point-to-plane distance as evaluation metrics.
Specifically, the absolute angular difference between the normals to the automatic and ground truth planes is computed, and the distance from the center of the ground truth view (image) to the automatic plane is measured.
For both metrics, smaller is better.

\begin{table}[!t]
  \captionv{16.5}{}{Sequence parameters used for the CMR image acquisition.
  All images were acquired with the 1.5T Siemens Symphony model.\label{tab:protocol}}
  \begin{center}
  \setlength{\tabcolsep}{.8mm}
  \begin{adjustbox}{width=\textwidth}
  \begin{threeparttable}
  \begin{tabular}{rcccccccc}
    \hline
    % after \\: \hline or \cline{col1-col2} \cline{col3-col4} ...
    \multicolumn{1}{c}{View} & Axial & p2C & p4C & pSA & 2C & 3C & 4C & SAX \\
    \hline
    \multicolumn{1}{c}{\multirow{ 2}{*}{Sequence}} & HASTE & \multirow{ 2}{*}{TRUFI} & \multirow{ 2}{*}{TRUFI} & \multirow{ 2}{*}{TRUFI} & TrueFISP & TrueFISP & TrueFISP & TrueFISP \\
    & dark blood & & & & cine retro & cine retro & cine retro & cine retro \\
    \hline
    \multicolumn{1}{l}{Flip angle ($^\circ$)} \\
    \cline{1-1}
    Mode/Mean\textsuperscript{a} & 160 (100\%) & 65 (100\%) & 65 (100\%) & 65 (99\%) & 73 (83\%) & 73 (86\%) & 74 (79\%) & 73 (84\%) \\
    Range & - & - & - & 64, 65 & 49--79 & 49--79 & 49--80 & 53--79 \\
    \hline
    \multicolumn{1}{l}{Rows (pixel)} \\
    \cline{1-1}
    Mode/Mean\textsuperscript{a} & 192 (88\%) & 192 (100\%) & 160 (85\%) & 192 (99\%) & 192 (100\%) & 170 & 155 & 192 (98\%) \\
    Range & 176--224 & - & 128--192 & 160, 176, 192 & - & 132--192 & 132--192 & 156, 162, 192 \\
    \hline
    \multicolumn{1}{l}{Columns (pixel)} \\
    \cline{1-1}
    Mode/Mean\textsuperscript{a} & 256 (100\%) & 176 (73\%) & 192 (97\%) & 176 (82\%) & 156 & 173 & 192 (98\%) & 161 \\
    Range & - & 144--192 & 160, 176, 192 & 144--192 & 144--180 & 132--192 & 144--192 & 144--192 \\
    \hline
    \multicolumn{1}{l}{Pixel spacing\textsuperscript{b} (mm)} \\
    \cline{1-1}
    Mode/Mean\textsuperscript{a} & 1.34 & 1.98 (66\%) & 1.77 (90\%) & 1.88 (95\%) & 1.80 & 1.67 (72\%) & 1.72 & 1.85 \\
    Range & 1.17--1.56 & 1.67--2.29 & 1.56--2.08 & 1.77--1.98 & 1.51--2.08 & 1.46--2.08 & 1.56--1.93 & 1.56--2.08 \\
    \hline
    \multicolumn{1}{l}{Slice thickness (mm)} \\
    \cline{1-1}
    Mode & 6 (100\%) & 6 (100\%) & 6 (100\%) & 6 (88\%) & 7 (96\%) & 7 (96\%) & 7 (96\%) & 7 (96\%) \\
    Range & - & - & - & 6, 6.5, 7 & 7, 8 & 6, 7, 8 & 7, 8 & 7, 8 \\
    \hline
    \multicolumn{1}{l}{Repetition time (ms)} \\
    \cline{1-1}
    Mode/Mean\textsuperscript{a} & 731 & 437 & 437 (93\%) & 426 (96\%) & 42.22 & 43.16 & 42.66 & 41.94 \\
    Range & 470--1000 & 400--798 & 371--840 & 423--857 & 40.18--63.8 & 40.46--51.04 & 41.02--63.2 & 39.9--45.08 \\
    \hline
    \multicolumn{1}{l}{Echo time (ms)} \\
    \cline{1-1}
    Mode/Mean\textsuperscript{a} & 28 (99\%) & 1.24 (89\%) & 1.27 (94\%) & 1.26 (95\%) & 1.28 & 1.31 & 1.29 & 1.27 \\
    Range & 28, 38 & 1.19--1.30 & 1.21--1.34 & 1.23--1.29 & 1.21--1.36 & 1.21--1.39 & 1.24--1.34 & 1.21--1.37 \\
    \hline
    \multicolumn{1}{l}{Number of slices} \\
    \cline{1-1}
    Mode/Mean\textsuperscript{a} & 30 (89\%) & 1 (100\%) & 1 (100\%) & 8 (83\%) & 1 (100\%) & 1 (100\%) & 1 (100\%) & 10.7 \\
    Range & 30--37 & - & - & 6--10 & - & - & - & 3--14 \\
    \hline
    \multicolumn{1}{l}{Slice interval\textsuperscript{c} (mm)} \\
    \cline{1-1}
    Mode/Mean\textsuperscript{a} & 6 (99\%) & - & - & 12 (85\%) & - & - & - & 10 (99\%) \\
    Range & 6, 6.6 & - & - & 10.8--15 & - & - & - & 5, 9.95, 10 \\
    \hline
  \end{tabular}
  \begin{tablenotes}
    \item[a] The mode is shown if it occupies the vast majority of the dataset (with the ratio in parentheses), otherwise{\color{red},} the mean is shown.% (with the standard deviation in parentheses)
    \item[b] Pixels are isotropic.
    \item[c] Measured from center-to-center of each slice along the common normal to the stack of slices.\\
    Abbreviations: HASTE: half-Fourier acquisition single-shot turbo spin-echo;
    TRUFI: true fast imaging with steady-state free precession;
    TrueFISP: true fast imaging with steady-state free precession.
  \end{tablenotes}
  \end{threeparttable}
  \end{adjustbox}
  \end{center}
\end{table}

\subsection{Implementation}
The PyTorch framework (1.4.0) is used for all experiments. %\cite{steiner2019pytorch}
%The commonly used encoder-decoder structure U-Net\cite{ronneberger2015u} is used as our backbone network.
A model is trained for each of the four localizers (axial, p2C, p4C, and pSA).
For stacked localizers (the axial and pSA), all slices of a patient are treated as a {\color{red}mini-batch} with the original image size as {\color{red}the} input size.
For the others, we use a {\color{red}mini-batch} size of eight images, whose sizes are unified according to data statistics by cropping or padding, where appropriate, for training;
specifically, the p2C and p4C localizers are unified to 192$\times$176 (rows by columns) pixels and 160$\times$192 pixels, respectively.
The Adam\cite{kingma2014adam} optimizer is used with a weight decay of 0.0005.
The learning rate is initialized to 0.001 and halved when the loss does not decrease for 10 consecutive epochs.
{\color{purple}We train the models until convergence.}
%The exact numbers of training epochs vary with the specific localizer, value of $\sigma$, and number of hourglasses, and are determined with the validation set (range 75--250).
%{\color{yellow}The numbers of epochs for learning rate adjustments and total training periods determined on the validation set are reused to train the final testing models, where the training and validation sets are combined for training.}
Online data augmentation including random size scaling ($[0.9, 1.1]$), rotation ($[-10^\circ, 10^\circ]$), cropping, and flipping (left-right for the p2C, upside-down for the p4C, and none for stacked localizers) is conducted during training to alleviate overfitting.
For simple preprocessing, the z-score standardization (mean subtraction followed by division by standard deviation) is performed per localizer per exam.
An NVIDIA Tesla P40 GPU is used for model training for the p2C, p4C, and pSA localizers, whereas four such GPUs are used for the axial.
For testing, a single Tesla P40 is used for all the localizers.

{\color{cyan}A U-Net implementation is employed as the baseline backbone network,\footnote{https://github.com/ShawnBIT/UNet-family} which modifies the original version\cite{ronneberger2015u} from three main aspects: (i) zero padding is used to maintain the spatial size, (ii) transposed convolutions\cite{zeiler2014visualizing} are used for feature upsampling, and (iii) batch normalization is incorporated for better training. %\cite{ioffe2015batch}
Convolutions (1$\times$1) without activation function are used to produce output for the regression tasks.
%{\color{yellow}How about halving filters?}
With the baseline backbone, we first study the impact of $\sigma$ in Eqn. (\ref{eq:heatmap}) %(Sect. \ref{sec:experiment:sigma})
and the efficacy of multi-view aggregation. %(Sect. \ref{sec:experiment:multi}).
Next, we investigate the optimal number of hourglasses for the
%stacked hourglass network
SHG-Net. %(Sect. \ref{sec:experiment:hourglass}).
Specifically, the basic building block is a dwindled U-Net of 1/2 depth of the baseline network, yielded by halving the number of convolution layers in each feature scale.
Thus, the two-hourglass SHG-Net is about the same total depth (in {\color{red}term of the} number of layers) and size (in {\color{red}terms of the} number of parameters) as the baseline backbone for a fair comparison, %between two models of comparable sizes,
as similarly done by Newell et al.\cite{newell2016stacked}
Lastly, we train a final model with the optimal configuration determined by the empirical studies above for performance evaluation.
Our implementation available at https://github.com/wd111624/CMR\_plan.}

\subsection{Empirical Study on System Design}
In this section, we empirically determine the optimal design of the proposed system on the validation set.
Concretely, we first study the impact of varying $\sigma$
%the Gaussian kernel size
in Eqn. (\ref{eq:heatmap}) on performance, and the efficacy of the proposed multi-view aggregation, both using the baseline
%U-Net\cite{ronneberger2015u} as the
backbone network.
On top of that, we then determine the number of stacked hourglasses.
% with deep supervision.
Lastly, we also verify the validation-set-preferred design of multi-view aggregation and hourglass stacking on the test set.

\noindent\textbf{Impact of Hyperparameter $\sigma$.}\label{sec:experiment:sigma}
We first study the impact of
%Gaussian kernel size
$\sigma$ in Eqn. (\ref{eq:heatmap}) by setting it to different values and comparing the validation performance.
%{\color{yellow}What would larger and smaller $\sigma$ do physically?}
Specifically, $\sigma$ is defined as a ratio/multiple of the slice thickness $t$ of the target view: $\sigma=\alpha t$, where $\alpha\in \{0.25, 0.5, 1, 2\}$.
The results are shown in Table \ref{tab:kernel_size}.
Based on a comprehensive consideration of both metrics on all the four target planes, we choose $\sigma=0.5t$ for subsequent experiments.

\begin{table}[!t]
\captionv{16.5}{}{Impact of the hyperparameter $\sigma$ on plane prescription accuracy on the validation set.
$\sigma$ is defined as a ratio/multiple of the slice thickness $t$: $\sigma=\alpha t$ with $\alpha$ being a factor.
Data format: mean $\pm$ standard deviation.\label{tab:kernel_size}}%All relevant source views are involved for the prescription of a target plane.
\begin{center}
%\setlength{\tabcolsep}{.7mm}
%\begin{adjustbox}{width=\textwidth}
%\begin{tabular}{cccccccccccc}
%\hline
%         & \multicolumn{5}{c}{Normal deviation ($^{\circ}$)}                                      &  & \multicolumn{5}{c}{Point-to-plane distance (mm)}                                       \\ \cline{2-6} \cline{8-12}
%$\sigma$ & 2C            & 3C            & 4C            & SAX           & Mean                   &  & 2C            & 3C            & 4C            & SAX           & Mean                   \\ \hline
%$0.25t$  & 4.71$\pm$2.44 & 7.14$\pm$4.18 & 7.33$\pm$5.16 & 8.40$\pm$5.10 & 6.89$\pm$1.56          &  & 2.66$\pm$2.02 & 2.44$\pm$2.47 & 3.76$\pm$3.03 & 3.11$\pm$2.37 & \textbf{2.99}$\pm$0.58 \\
%$0.5t$   & 4.86$\pm$2.95 & 7.18$\pm$4.62 & 6.85$\pm$4.54 & 7.97$\pm$4.61 & \textbf{6.71}$\pm$1.32 &  & 2.78$\pm$1.92 & 2.76$\pm$2.70 & 3.76$\pm$3.09 & 3.07$\pm$2.45 & 3.09$\pm$0.47          \\
%$1.0t$   & 5.82$\pm$3.46 & 7.42$\pm$4.90 & 7.09$\pm$3.96 & 8.04$\pm$5.91 & 7.09$\pm$0.94          &  & 3.47$\pm$2.51 & 3.32$\pm$3.35 & 3.63$\pm$2.83 & 3.77$\pm$2.87 & 3.55$\pm$0.20          \\
%$2.0t$   & 6.29$\pm$4.14 & 7.97$\pm$5.29 & 7.50$\pm$4.98 & 8.21$\pm$4.53 & 7.49$\pm$0.85          &  & 4.88$\pm$3.23 & 3.87$\pm$3.45 & 4.59$\pm$3.56 & 4.38$\pm$2.57 & 4.43$\pm$0.43          \\ \hline
%\end{tabular}
%\end{adjustbox}
\setlength{\tabcolsep}{3.7mm}
%\begin{adjustbox}{width=.73\textwidth}
\begin{tabular}{cccccc}
\hline
         & 2C            & 3C            & 4C            & SAX           & Mean                   \\ \cline{2-6}
$\sigma$ & \multicolumn{5}{c}{\emph{Normal deviation} ($^{\circ}$)}                                      \\ \hline
$0.25t$  & 4.71$\pm$2.44 & 7.14$\pm$4.18 & 7.33$\pm$5.16 & 8.40$\pm$5.10 & \underline{6.89}$\pm$1.56          \\
%\rowcolor[HTML]{EFEFEF}
$0.5t$   & 4.86$\pm$2.95 & 7.18$\pm$4.62 & 6.85$\pm$4.54 & 7.97$\pm$4.61 & \textbf{6.71}$\pm$1.32 \\
$1.0t$   & 5.82$\pm$3.46{\color{blue}*} & 7.42$\pm$4.90 & 7.09$\pm$3.96 & 8.04$\pm$5.91 & 7.09$\pm$0.94          \\
$2.0t$   & 6.29$\pm$4.14{\color{blue}**} & 7.97$\pm$5.29 & 7.50$\pm$4.98 & 8.21$\pm$4.53 & 7.49$\pm$0.85{\color{blue}*} \\ \hline
$\sigma$ & \multicolumn{5}{c}{\emph{Point-to-plane distance} (mm)}                                       \\ \hline
$0.25t$  & 2.66$\pm$2.02 & 2.44$\pm$2.47 & 3.76$\pm$3.03 & 3.11$\pm$2.37 & \textbf{2.99}$\pm$0.58 \\
%\rowcolor[HTML]{EFEFEF}
$0.5t$   & 2.78$\pm$1.92 & 2.76$\pm$2.70 & 3.76$\pm$3.09 & 3.07$\pm$2.45 & \underline{3.09}$\pm$0.47          \\
$1.0t$   & 3.47$\pm$2.51 & 3.32$\pm$3.35 & 3.63$\pm$2.83 & 3.77$\pm$2.87{\color{blue}*} & 3.55$\pm$0.20{\color{blue}*}          \\
$2.0t$   & 4.88$\pm$3.23{\color{blue}***} & 3.87$\pm$3.45{\color{blue}*} & 4.59$\pm$3.56{\color{blue}*} & 4.38$\pm$2.57{\color{blue}**} & 4.43$\pm$0.43{\color{blue}**} \\
\hline
\multicolumn{6}{l}{{\emph{Note:} *, **, *** represent $p\leq0.05$, $0.01$, and $0.001$, respectively, for comparisons}}\\
\multicolumn{6}{l}{{between $\sigma=0.5t$ and alternatives using {\color{red}\emph{t}-test}.}}\\
\end{tabular}
%\end{adjustbox}
\end{center}
\end{table}

\noindent\textbf{Efficacy of Multi-View Aggregation.}\label{sec:experiment:multi}
Next, we investigate the impact of the proposed multi-view planning on plane prescription accuracy.
{\color{cyan}As introduced in Sect. \ref{sec:preliminary:multiview}, the prescription lines in two non-coinciding source views are sufficient to define a target plane.
Therefore, it is feasible to prescribe some standard CMR planes from a subset of available localizers in our protocol: 3C plane from the two pLA (p2C and p4C) localizers, and 2C, 3C, and 4C planes from the stack of pSA localizers.\footnote{\color{yellow}Here we exclude the extreme cases of single-view prescription, whose presumption of perpendicularity contradicts with our experiment data where the standard planes were prescribed based on multiple localizers.}}
%In theory, it is possible to define a target plane with only a subset of available source views, as the intersecting lines with two non-coinciding planes are sufficient to define a definite plane.
%We examine such cases in our scenario and present the results in Table \ref{tab:multiview}.
%(e.g., the stack of SAX localizer alone can be used to determine any of the standard LAX views, and the 2C and 4C LAX localizers can be used jointly to prescribe the 3C LAX cine plane without the SAX localizers)
The results in Table \ref{tab:multiview} show that despite being theoretically feasible, prescribing the standard planes using a subset of localizers leads to generally inferior performance, suggesting the importance of multi-view planning.
This is consistent with the clinical practice, where the operators consider multiple localizers to prescribe a standard plane.
Notably, when using only the two pLA localizers to prescribe the 3C plane, the performance collapses with mean normal deviations of 41.88$^\circ$ and 28.12$^\circ$ on the validation and test sets, respectively.
We speculate that this is because the key landmark (i.e., the aortic valve) that defines the 3C plane is only visible in the stack of pSA localizers.

\begin{table}[!t]
\captionv{16.5}{}{Impact of multi-view planning on prescription accuracy on the validation (left) and test (right) sets (with $\sigma=0.5t$);
note that `pLA' includes both p2C and p4C localizers.
`-' indicates that a specific plane cannot be prescribed solely from the corresponding subset of localizers in our imaging protocol.
Data format: mean $\pm$ standard deviation.\label{tab:multiview}}
\begin{center}
%\setlength{\tabcolsep}{1.55mm}
%\begin{adjustbox}{width=.95\textwidth}
%\begin{tabular}{cccccccccc}
%\hline
%           & \multicolumn{4}{c}{Normal deviation ($^{\circ}$)}                        &  & \multicolumn{4}{c}{Point-to-plane distance (mm)}                       \\ \cline{2-5} \cline{7-10}
%Localizer  & 2C            & 3C              & 4C            & Mean                   &  & 2C            & 3C            & 4C            & Mean                   \\ \hline
%pLA        & -             & 41.88$\pm$29.98 & -             & NA                     &  & -             & 7.83$\pm$7.02 & -             & NA                     \\
%pSA        & 4.96$\pm$2.77 & 7.68$\pm$4.67   & 6.80$\pm$4.16 & 6.48$\pm$1.39          &  & 3.33$\pm$2.33 & 3.11$\pm$2.59 & 3.92$\pm$3.29 & 3.45$\pm$0.42          \\
%pLA \& pSA & 4.86$\pm$2.95 & 7.18$\pm$4.62   & 6.85$\pm$4.54 & \textbf{6.30}$\pm$1.26 &  & 2.78$\pm$1.92 & 2.76$\pm$2.70 & 3.76$\pm$3.09 & \textbf{3.10}$\pm$0.57 \\ \hline
%\end{tabular}
%\end{adjustbox}
\setlength{\tabcolsep}{.7mm}
\begin{adjustbox}{width=\columnwidth}
\begin{tabular}{cccccccccc}
\hline
 & \multicolumn{5}{l}{Validation set} & \multicolumn{4}{l}{Test set} \\
\cline{2-5}\cline{7-10}
 & 2C            & 3C              & 4C            & Mean & & {2C}            & {3C}              & {4C}            & {Mean}                  \\\hline
Localizer & \multicolumn{9}{c}{\emph{Normal deviation} ($^{\circ}$)}                        \\ \hline
pLA        & -             & 41.88$\pm$29.98*** & -             & NA & & {-}             & {34.86$\pm$28.21***} & {-}             & {NA}                     \\
pSA        & 4.96$\pm$2.77 & 7.68$\pm$4.67   & 6.80$\pm$4.16 & 6.48$\pm$1.39 & & {4.63$\pm$3.07} & {6.90$\pm$4.12}  & {5.96$\pm$2.55} & {5.83$\pm$1.14} \\
%\rowcolor[HTML]{EFEFEF}
pLA \& pSA & 4.86$\pm$2.95 & 7.18$\pm$4.62   & 6.85$\pm$4.54 & \textbf{6.30}$\pm$1.26 & & {4.81$\pm$2.97} & {6.58$\pm$3.94} & {5.75$\pm$2.82} & {\textbf{5.71}$\pm$0.88} \\ \hline
Localizer  & \multicolumn{9}{c}{\emph{Point-to-plane distance} (mm)} \\ \hline
pLA        & -             & 7.83$\pm$7.02***   & -             & NA & & {-} & {6.18$\pm$4.27***} & {-} & {NA}              \\
pSA        & 3.33$\pm$2.33* & 3.11$\pm$2.59   & 3.92$\pm$3.29 & 3.45$\pm$0.42{*} & & {3.08$\pm$2.66*}  & {3.46$\pm$2.87} & {4.13$\pm$2.36} & {3.56$\pm$0.53*} \\
%\rowcolor[HTML]{EFEFEF}
pLA \& pSA & 2.78$\pm$1.92 & 2.76$\pm$2.70   & 3.76$\pm$3.09 & \textbf{3.10}$\pm$0.57 & & {2.61$\pm$2.46} & {2.99$\pm$2.57} & {3.96$\pm$2.49} & {\textbf{3.19}$\pm$0.70} \\ \hline
%\multicolumn{10}{l}{{\color{blue}\emph{Note:} \textsuperscript{a} The test set results are obtained by our finalized model.}}\\
\multicolumn{10}{l}{{\emph{Note:} *, **, *** represent $p\leq0.05$, $0.01$, and $0.001$, respectively, for comparisons between multi-view planning}}\\
\multicolumn{10}{l}{{(pLA \& pSA) and alternatives using {\color{red}\emph{t}-test}.}}\\
\multicolumn{5}{l}{Abbreviation: NA---not applicable.}
\end{tabular}
\end{adjustbox}
\end{center}
\end{table}

\begin{figure}[t]
\begin{center}
  \includegraphics[width=16cm]{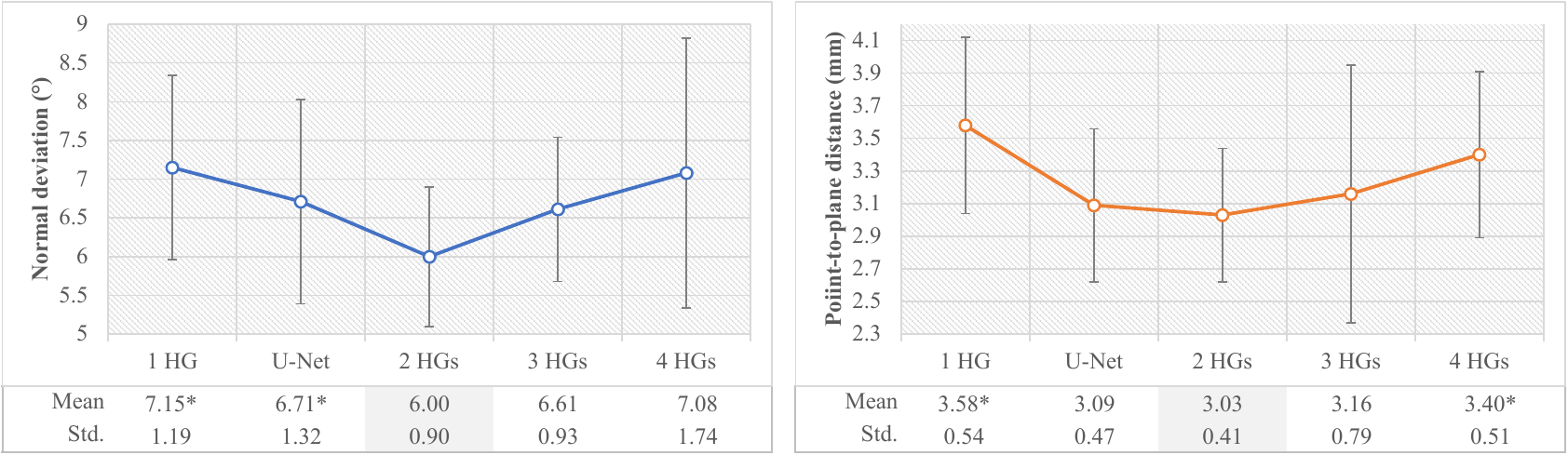}
  \captionv{16.5}{}{\color{cyan}Mean prescription performance and standard deviations (std.; averaged across all the standard CMR planes, i.e., 2C, 3C, 4C, and SAX) on the validation set, as the number of stacked hourglasses (HGs) varies.
  The performance of using the baseline backbone (U-Net) is also compared, which is about the same depth (in {\color{red}terms of the} number of layers) and size (in {\color{red}terms of the} number of parameters) as two HGs.
  \emph{Note:} * represent $p\leq0.05$ for comparisons between 2 HGs and alternatives using {\color{red}\emph{t}-test}.\label{fig:num_HGs}}
\end{center}
\end{figure}

\noindent\textbf{Number of Stacked Hourglasses.}\label{sec:experiment:hourglass}
Based on the optimal $\sigma$ value and using the multi-view aggregation, we then determine the optimal number of hourglasses to stack.
Specifically, we vary the number of hourglasses from {\color{cyan}one} to four {\color{red}and} compare the resulting mean prescription accuracy averaged across the standard LAX and SAX planes.
As shown in Fig. \ref{fig:num_HGs}, while three hourglasses yield modestly inferior performance to two, four hourglasses lead to further apparent performance degradation in both evaluation metrics.
%(which is the reason why we do not experiment with even more hourglasses)
Meanwhile, compared to the baseline backbone U-Net,
%(\emph{i.e.}, U-Net about the same depth and size)
two hourglasses are appreciably superior in normal deviation (by $\sim$0.7$^\circ$) and marginally superior in {\color{red}the} point-to-plane distance (by less than 0.1 mm), respectively, {\color{cyan}and also more stable for prescribing different planes as indicated by the smaller standard deviations.
It is worth noting that as the two-hourglass SHG-Net is about the same total depth and size as the baseline network, the improvements are not due to more network layers or parameters.
Lastly, one hourglass cannot produce as good performance, probably due to  its small model capacity.}
Therefore, we choose the SHG-Net with two hourglasses for subsequent performance evaluation on the test set.

Table \ref{tab:HG_testset_cmp} shows the performance comparison between the two-hourglass SHG-Net and the baseline backbone using U-Net---which {\color{red}is our} preliminary version\cite{wei2021training}---on the test set.
As we can see, the SHG-Net excels in almost all paired comparisons (i.e., 9 of 10). Especially, it outperforms the baseline in the mean normal deviation and point-to-plane distance averaged across the standard CMR planes by $\sim$5\% and $\sim$10\%, respectively.
These improvements verify the positive effect of the iterative refinement brought by the stacked hourglass architecture.

\begin{table}[t]
  \captionv{16.5}{}{Test set performance comparison between the proposed two-hourglass SHG-Net and the baseline backbone\cite{wei2021training} using U-Net.
  \label{tab:HG_testset_cmp}}
  \begin{center}
  %\setlength{\tabcolsep}{1.2mm}
  %\begin{adjustbox}{width=\textwidth}
  %\begin{threeparttable}
  \begin{tabular}{cccccc}
\hline
        & 2C       & 3C      & 4C     & SAX     & Mean     \\ \cline{2-6}
Network & \multicolumn{5}{c}{\emph{Normal deviation} ($^\circ$)}             \\ \hline
SHG-Net & \textbf{4.77}$\pm$3.06 & \textbf{6.21}$\pm$3.87 & 6.28$\pm$3.98 & \textbf{5.45}$\pm$3.51 & \textbf{5.68}$\pm$0.71 \\
Baseline\cite{wei2021training}   & 4.97$\pm$4.00 & 6.84$\pm$4.16 & \textbf{5.84}$\pm$3.19 & 6.28$\pm$3.48* & 5.98$\pm$0.79 \\ \hline
        & \multicolumn{5}{c}{\emph{Point-to-plane distance} (mm)} \\ \hline
SHG-Net & \textbf{2.35}$\pm$1.92 & \textbf{2.90}$\pm$2.60 & \textbf{3.34}$\pm$2.24 & \textbf{3.88}$\pm$3.38 & \textbf{3.12}$\pm$0.65 \\
Baseline\cite{wei2021training}   & 2.68$\pm$2.34 & 3.44$\pm$2.37** & 3.61$\pm$2.63 & 4.18$\pm$3.15 & 3.48$\pm$0.62** \\ \hline
\multicolumn{6}{l}{{\emph{Note:} * and ** represent $p\leq0.05$ and $0.01$, respectively, for {\color{red}\emph{t}-test}.}}\\
\end{tabular}
  %\end{threeparttable}
  %\end{adjustbox}
  \end{center}
\end{table}

\subsection{System-Level Evaluation}%Evaluation on Test-Set and Comparison to Existing Methods
This section reports the evaluation performance on the test set, using the optimal system design determined above.
Prescription of the standard CMR planes is first evaluated and compared to that of existing works, followed by prescription of the pseudo-view localizers---which is the first step from body-oriented imaging planes to the {\color{red}cardiac-anatomy-oriented} ones in CMR, but absent in existing literature.

\begin{table}[t]
  \captionv{16.5}{}{Test set evaluation results (Proposed\textsuperscript{ test}) and comparison with previous works.
  %Bold and underlined numbers indicate the best and second best results per column and per evaluation metric, respectively.
  The results {\color{red}of} the training and validation sets (Proposed\textsuperscript{ train+val}) are also presented for reference.
  -: not reported.
  %NA: not applicable.
  %*: significance at 0.05 level for pairwise comparison to the proposed method (Bonferroni correction applied where appropriate;
  %since different works used different data, independent samples \emph{t}-test is employed.
  Data format: mean $\pm$ standard deviation.
  %{\color{cyan}($p$ value), where the $p$ value is for pairwise comparison to the proposed method using independent samples (for view-specific comparisons) or paired (for cross-view mean comparisons) \emph{t}-test}.
  \label{tab:cmp}}%results reported in the literature cannot be directly compared but provide reference values
  \begin{center}
  \setlength{\tabcolsep}{1.2mm}
  \begin{adjustbox}{width=\textwidth}
  %\begin{threeparttable}
  \begin{tabular}{rccccc}
    \hline
    % after \\: \hline or \cline{col1-col2} \cline{col3-col4} ...
    %\multicolumn{7}{c}{\emph{Normal deviation} ($^\circ$)} \\
    %\Xhline{.8pt}
     & 2C & 3C & 4C & SAX & Mean \\
    \cline{2-6}%\Xhline{1pt}
    Methods & \multicolumn{5}{c}{\emph{Normal deviation} ($^\circ$)} \\
    \hline
    Lu et al.\cite{lu2011automatic} & 18.9$\pm$21.0*** & 12.3$\pm$11.0*** & 17.6$\pm$19.2*** & 8.6$\pm$9.7 & 14.35$\pm$4.78* \\
    Alansary et al.\cite{alansary2018automatic} & - & - & 8.72$\pm$7.44 & - & NA \\
    Frick et al.\cite{frick2011fully} & \underline{7.1}$\pm$3.6** & 9.1$\pm$6.3* & 7.7$\pm$6.1 & 6.7$\pm$3.6 & 7.65$\pm$1.05* \\
    Blansit et al.\cite{blansit2019deep} & 8.00$\pm$6.03** & \underline{7.19}$\pm$4.97 & \textbf{5.49}$\pm$5.06 & \underline{5.56}$\pm$4.60 & \underline{6.56}$\pm$1.24 \\
    %\cite{wei2021training} & \underline{4.97}$\pm$4.00 & \underline{6.84}$\pm$4.16 & \underline{5.84}$\pm$3.19 & 6.28$\pm$3.48 & \underline{5.98}$\pm$0.79 \\
    %\rowcolor[HTML]{EFEFEF}
    Proposed\textsuperscript{ test} & \textbf{4.77}$\pm$3.06 & \textbf{6.21}$\pm$3.87 & \underline{6.28}$\pm$3.98 & \textbf{5.45}$\pm$3.51 & \textbf{5.68}$\pm$0.71 \\
    \hdashline
    {Proposed\textsuperscript{ train+val}} & {3.42$\pm$2.46} & {5.77$\pm$4.52} & {4.87$\pm$3.38} & {4.94$\pm$3.22} & {4.75$\pm$0.97} \\
    \hline%\Xhline{1pt}
    Methods & \multicolumn{5}{c}{\emph{Point-to-plane distance} (mm)} \\
    \hline
    Lu et al.\cite{lu2011automatic} & {6.6}$\pm$8.8** & {4.6}$\pm$7.7 & 5.7$\pm$8.5 & 13.3$\pm$16.7*** & 7.55$\pm$3.92 \\
    Alansary et al.\cite{alansary2018automatic} & - & - & \underline{5.07}$\pm$3.33** & - & NA \\
    %\cite{wei2021training} & \underline{2.68}$\pm$2.34 & \underline{3.44}$\pm$2.37 & \underline{3.61}$\pm$2.63 & \underline{4.18}$\pm$3.15 & \underline{3.48}$\pm$0.62 \\
    %\rowcolor[HTML]{EFEFEF}
    Proposed\textsuperscript{ test} & \textbf{2.35}$\pm$1.92 & \textbf{2.90}$\pm$2.60 & \textbf{3.34}$\pm$2.24 & \textbf{3.88}$\pm$3.38 & \textbf{3.12}$\pm$0.65 \\
    \hdashline
    {Proposed\textsuperscript{ train+val}} & {2.09$\pm$2.19} & {2.67$\pm$3.04} & {2.74$\pm$2.46} & {2.87$\pm$2.24} & {2.59$\pm$0.35} \\
    \hline
    \multicolumn{6}{l}{\emph{Note}: *, **, *** represent $p\leq0.05$, $0.01$, and $0.001$, respectively, for comparisons between}\\
    \multicolumn{6}{l}{our test set results (Proposed\textsuperscript{ test}) and previous works using {\color{red}$t$-test}.}\\
    \multicolumn{6}{l}{Abbreviation: NA---not applicable.}
  \end{tabular}
  %\end{threeparttable}
  \end{adjustbox}
  \end{center}
\end{table}

\noindent\textbf{Prescription of Standard CMR Planes.}
We now evaluate our proposed system on the held-out test set, and compare the performance with existing approaches\cite{alansary2018automatic,blansit2019deep,frick2011fully,lu2011automatic}.\footnote{\color{cyan}Caution: due to various objective constraints (inaccessibility to valid implementation, lack of manual landmark annotation, or lack of appropriate data), the studies do not share common reference datasets.}
We also provide the evaluation results on the training and validation sets for reference {\color{red}purposes}.
The results are charted in Table \ref{tab:cmp}.
Above all, our method achieves the best mean performance averaged over the four standard CMR planes in both evaluation metrics (mean normal deviation: 5.68$\pm$0.71$^\circ$ {\color{red}and} mean point-to-plane distance: 3.12$\pm$0.65~mm).
The second best method\cite{blansit2019deep} in terms of the mean normal deviation is $\sim$15\% shy  (6.56$\pm$1.24$^\circ$), whereas the only {\color{red}existing} method that computed the point-to-plane distances for all the four standard planes reported a mean distance of 7.55$\pm$3.92 mm, which is $\sim$142\% higher than ours.
Further {\color{red}per-plane} analysis indicates that our method is also competent for each plane: it yields the best point-to-plane distances for {\color{red}all four} standard planes, the lowest normal deviations for three (i.e., 2C, 3C, and SAX) of the four, and the second lowest normal deviation for the remaining 4C plane.
{\color{cyan}Specifically, our mean normal deviation (5.45$\pm$3.51$^\circ$) for the SAX plane is close to the inter-operator variation (4.99$\pm$2.17$^\circ$) reported by Danilouchkine et al.\cite{danilouchkine2005operator}
%Danilouchkine et al.\cite{danilouchkine2005operator} reported mean SAX normal deviations of 2.67$\pm$1.5$^\circ$ and 4.99$\pm$2.17$^\circ$ for the intra- and inter-operator variations, respectively.
}

\begin{table}[t]
\captionv{16.5}{}{Test set evaluation results for the pseudo-view localizers.
  Data format: mean $\pm$ standard deviation.
  The results {\color{red}of} the training and validation sets (train+val) are also presented for reference.\label{tab:loc_results}}
\begin{center}
\color{purple}
\setlength{\tabcolsep}{1.8mm}
%\begin{adjustbox}{width=.9\textwidth}
\begin{tabular}{cccccc}
\hline
Dataset & Metric & p2C           & p4C           & pSA           & Mean          \\ \hline
Test & Normal deviation ($^\circ$)  & 6.23$\pm$3.97 & 5.64$\pm$4.76 & 5.64$\pm$3.84 & 5.84$\pm$0.34 \\
& Point-to-plane distance (mm) & 3.80$\pm$2.99 & 4.15$\pm$3.06 & 3.91$\pm$1.43 & 3.95$\pm$0.18 \\ \hdashline
{Train+val} & {Normal deviation ($^\circ$)} & {5.26$\pm$3.29} & {4.97$\pm$5.95} & {4.63$\pm$3.52} & {4.95$\pm$0.32} \\
& {Point-to-plane distance (mm)} & {3.45$\pm$2.83} & {2.55$\pm$2.28} & {3.17$\pm$1.85} & {3.06$\pm$0.46} \\ \hline
\end{tabular}
%\end{adjustbox}
\end{center}
\end{table}

\noindent\textbf{Prescription of Pseudo-View Localizers.}\label{sec:exp:locs}
Last but not least, our method can also successfully prescribe the pseudo-view localizers
%given the axial scout
(Figs. \ref{fig:protocol}(a) and \ref{fig:protocol}(b)).
{\color{cyan}For prescribing the p2C plane from the stack of dark-blood axial slices (Fig. \ref{fig:protocol}(a)), we train another regression model from scratch with the same settings as training on the stack of pSA slices.
As to prescribing the p4C and pSA planes from the p2C localizer (Fig. \ref{fig:protocol}(b)), we fine-tune the model trained for prescribing the standard planes for five epochs with the learning rate of 0.0002, given the considerable similarity between the two tasks.}
%{\color{purple}Note that for the prescription of the localizers we fine-tune the CNN models trained for the standard views for five epochs with the learning rate of 0.0002, considering the inadequacy of the available training data due to the diverging alternative protocols.}
Results in Table \ref{tab:loc_results} show that the performance is comparable to that of prescribing the standard CMR planes, with the mean (over the localizers) normal deviation of {\color{purple}5.84$\pm$0.34$^\circ$} and point-to-plane distance of {\color{purple}3.95$\pm$0.18 mm on the held-out test set}.

\begin{figure}[t]
\begin{center}
    \includegraphics[width=16cm,trim=0 0 0 0,clip]{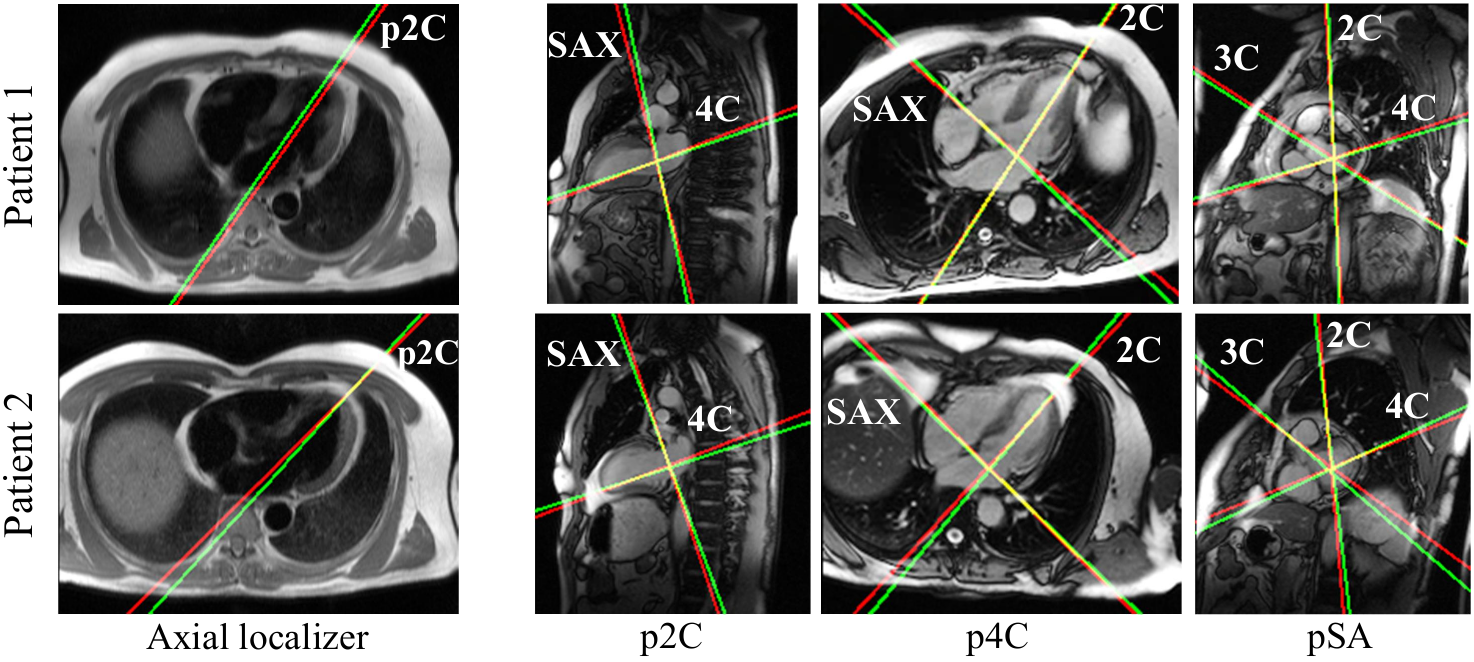}
    \captionv{16.5}{}{Example CMR plane prescription results by the proposed system for two patients.
    Green: ground truth; red: automatic prescription; yellow: overlap (online version only).
    Due to the closeness (sometimes coincidence) of the 3C plane with the 2C and 4C planes in the p4C and p2C localizers, the 3C plane is visualized only in the pSA localizer.\label{fig:results}}
\end{center}
\end{figure}

\begin{figure}[t]
\begin{center}
  \includegraphics[width=16cm]{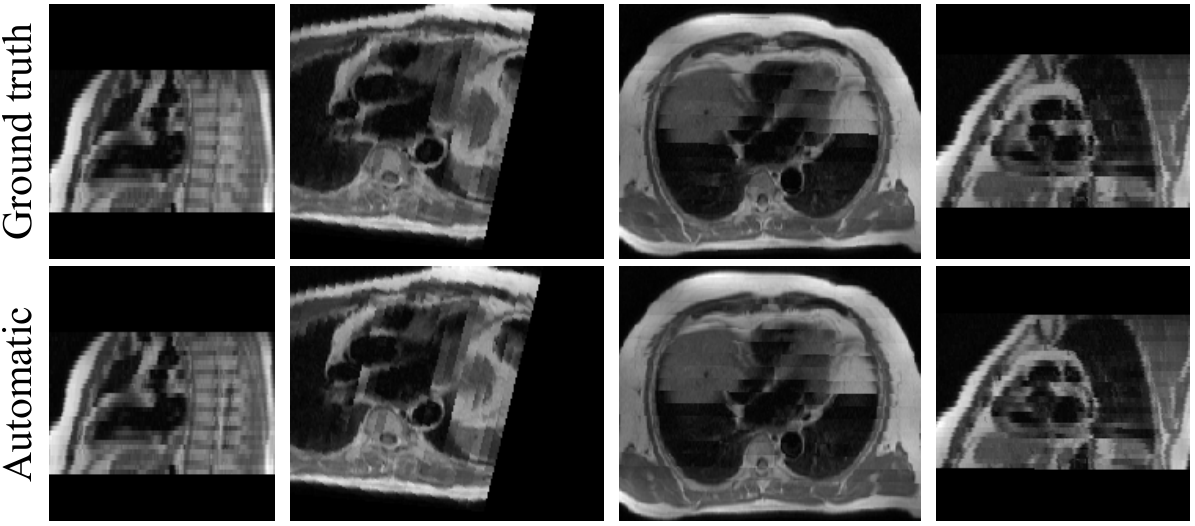}
  \captionv{16.5}{}{\color{cyan}Left to right: 2C, 3C, 4C, and SAX standard-view images resampled from the dark-blood axial localizer according to automatic and manual ground truth prescriptions, for Patient 1 in Fig. \ref{fig:results}.\label{fig:resample}}
\end{center}
\end{figure}

\noindent\textbf{Result Visualization.}
Some examples of the automatic prescription results are visualized in Fig.~\ref{fig:results}, {\color{cyan}in the form of prescription lines within the localizers.}
We can see that the CMR planes prescribed by the proposed system are fairly close to the ground truth prescribed by the technologists, especially within the hearts.
The visualizations suggest that the CNN models {\color{red}have} learned knowledge about {\color{red}cardiac} anatomy, despite being indirectly supervised by the spatial relationship between intersecting views.
{\color{cyan}In addition, due to the retrospective nature of this study, we cannot actually acquire the standard views according to the automatic prescription.
Alternatively, we resample the stack of dark-blood axial localizers to ``generate'' the standard views, via nearest neighbor interpolation.
As shown in Fig. \ref{fig:resample}, the images resampled according to the automatic and manual ground truth prescriptions visualize cardiac structures similarly.}

\section{Discussion}\label{sec:discussion}
To the best of our knowledge, this work presented the first {\color{red}clinically} compatible system that demonstrated a complete automatic prescription pipeline for CMR imaging planes.
Prescribing the set of pseudo-view localizers, especially the p2C (vertical LAX) view, often is the first step from body-oriented imaging planes to the {\color{red}cardiac-anatomy-oriented} ones in CMR, yet it has not been previously demonstrated for any clinic-compatible system.
Blansit et al.\cite{blansit2019deep} assumed the p2C localizer was given as the starting point of their proposed automatic prescription system, yet they later discussed that achieving this first p2C prescription might require additional localization on an axial or sagittal stack.
% that do not rely on a volumetric scout.
In this work, besides demonstrating the state-of-the-art performance on prescribing standard CMR planes, we also demonstrated the prescription of the first p2C localizer given the stack of axial {\color{red}localizers}, and that of the subsequent p4C and pSA localizers---with a comparable performance.
These results not only demonstrated the competence and effectiveness of our proposed system in automatic CMR view planning {\color{red}but} also filled the critical gap towards clinical application of the automatic systems.

In practice, different healthcare institutions may practice quite different imaging protocols.
For example, while our data acquisition followed the standardized protocols\cite{kramer2020standardized} to adopt multi-view planning for the standard CMR views, Blansit et al.\cite{blansit2019deep} employed a simplified single-view prescription scheme.
%Mention a bit on different protocols: 2C bisect LV or 4C and 3C plane?
Thus, it was valuable that we also demonstrated the comprehensive applicability of our proposed system to versatile imaging protocols in this work.
Extended from our preliminary exploration\cite{wei2021training}, this work elaborated how the proposed system could also be applied to {\color{cyan}parallel-view and} single-view prescriptions, with minor adaption.
{\color{cyan}Given that the pseudo-view localizers in our imaging protocol were prescribed from a stack of parallel source views (for p2C; see Fig. \ref{fig:protocol}(a)) or a single source view (for p4C and pSA; see Fig. \ref{fig:protocol}(b)}), the consistent test-set evaluation results on them (Table \ref{tab:loc_results}) proved that our system could also work well for the alternative protocols.
In fact, since our proposed system did not make any assumption about the imaged anatomy, it has the potential to be widely applied to other complex clinical examinations that require precise prescription of anatomy-oriented imaging planes.
As shown in this work, as long as the plane prescription dependency and spatial information are available for the data, the proposed system should work.

The method proposed by Blansit et al.\cite{blansit2019deep} was most related to ours in that they both exploited the power of deep CNNs for the challenging task.
This method also achieved impressive performance---the {\color{red}second-best} cross-view mean normal deviation in Table \ref{tab:cmp}.
However, it required manual labeling of key anatomical landmarks, i.e., the apex and mitral valve in all LAX images, and the aortic, mitral, pulmonic, and tricuspid valves in SAX images.
The mass labeling process relied on relevant expertise {\color{red}and} can be time-consuming, exhausting, and error-prone.
In contrast, our method {\color{red}eliminated} this demanding requirement and directly learned from data, by mining the spatial relationship of the target plane to prescribe with its source views.
Although waived the need for manual annotation, our method surpassed {\color{red}the} performance of the previous work\cite{blansit2019deep} and produced even lower normal deviations.
Besides the different mechanisms of plane prescription (connecting detected key points versus directly regressing the prescription lines), we speculate that the difference in performance may also be explained by the fact that Blansit et al.\cite{blansit2019deep} built their system on single-view prescription, whereas we built ours on multi-view aggregation.

As suggested in the standardized protocols\cite{kramer2020standardized}, skilled operators often plan the CMR views by checking the locations of a candidate plane within multiple source views.
We mimicked this desirable behavior by prescribing the plane with the greatest collective response aggregated over its intersecting lines with all the source heatmaps.
{\color{red}The} efficacy of the proposed multi-view aggregation strategy was empirically validated by the results in Table \ref{tab:multiview}, where prescriptions with the multi-view aggregation were {\color{red}more accurate} than those with the pLA or pSA localizers alone.

Compared to the U-Net baseline of about the same depth and size, the stacked hourglass network (SHG-Net) delivered overall better validation performance (Fig. \ref{fig:num_HGs}).
In addition, the complete system proposed in this work also outperformed its preliminary version\cite{wei2021training}---which {\color{red}was} the U-Net baseline in this work---on the test set {\color{blue}(Table \ref{tab:HG_testset_cmp})}, reducing the cross-view mean normal deviation and point-to-plane distance by $\sim$5\% and $\sim$10\%, respectively.
These results suggested that the SHG-Net did effectively make use of the interplays between different target planes via the repeated top-down, bottom-up operations that took into account the features and predictions made by previous hourglasses, instead of relying on more learnable parameters.
{\color{red}When} we increased the number of the glasses to three and four, the performance dropped from slightly to {\color{red}large}.
%and when we increased that to four, the performance degraded .
We conjecture this unexpected phenomenon was a manifestation of overfitting due to the small sample size.

This work had limitations.
First, on one hand, a dataset of {\color{red}modest} size was used for experiments, with a total of 181 CMR exams from 99 patients. %compare to related works?
On the other hand, the dataset had a limited diversity, comprising only infarction patients from a single institute {\color{cyan}and imaged with a single scanner model and field strength}.
Consequently, we suspected that the effectiveness of the adopted SHG-Net with more hourglasses was negatively impacted by {\color{red}overfitting} the training data.
In the future, it is expected to validate the generalization of the proposed system on substantially larger, more diverse datasets containing both healthy subjects and other cardiac diseases, to push it towards real clinical application.
Second, due to the retrospective nature of this study, we could only evaluate the individual steps of our imaging protocol in Fig. \ref{fig:protocol}---with existing image data, instead of the integral pipeline where the pseudo-view localizers are firstly acquired according to our approach and then served as the basis for subsequent planning of standard CMR views.
To address the concern over potential error accumulation in {\color{red}multi-step planning} and prove genuine clinical usability, it is necessary to evaluate our method in a future prospective study.

In this work, we trained a separate model for each localizer instead of a general one for all, mainly for three factors:
(i) the large differences in appearance between the dark-blood axial and other bright-blood localizers (inter-sequence variation), {\color{red}and} anatomic landmarks between the pSA and pLA localizers (inter-view variation); 
(ii) the imbalance in numbers of the training images between single-slice (p2C and p4C) and multi-slice (axial and pSA) localizers; 
and (iii) the limited overall size of the dataset used. 
The combination of these factors raised the concern that the limited samples of each localizer would confuse the learning of a vanilla CNN targeted for simultaneous prediction of the prescription lines in all the localizers. 
In fact, our primitive attempts to train a single model with all the localizers together failed to predict reasonable prescription lines in any of them, leading to an eventual mean normal deviation of 53.66$^\circ$ averaged across the standard CMR planes. 
Alternatively, as an effective proof-of-concept study and the first of its kind, this work employed the straightforward localizer-specific model design and demonstrated that {\color{red}an} automatic CMR view planning system could be obtained from properly archived data—without requiring additional manual annotation, by mining the spatial relationship between the target planes and source views for training. 
Meanwhile, studies have shown that modern deep networks have the ability to learn and predict from a wide range of inputs, especially when the networks and datasets are both large and with effective designs to unify various tasks\cite{kirillov2023segment,brown2020language}.
It would be interesting to investigate such {\color{red}a} unified model design in our context, with a much larger collection of data in the future.

\section{Conclusion}\label{sec:conclusion}
In this work, we proposed a CNN-based, clinic-compatible system for automatic CMR view planning.
Our system was distinctive from a closely related work\cite{blansit2019deep} in that it eliminated the burdensome need for manual annotations, by mining the spatial relationship between CMR views.
Besides, the importance of the proposed multi-view aggregation---an {\color{red}analog} to the {\color{red}behavior} of human prescribers---was empirically validated.
Also, the interplay of the multiple target planes predicted in the same source view was effectively utilized in the stacked hourglass architecture for boosts in performance.
Experimental results showed that our system was superior to existing approaches including both conventional atlas-based and newer CNN-based ones, in prescription accuracy of four standard CMR planes.
%{\color{cyan}Noteworthily, the angular differences in standard SAX plane orientations between automatic prescriptions by our method and manual prescriptions were similar to the inter-observer variation\cite{danilouchkine2005operator}.}
In addition, we demonstrated accurate {\color{red}prescriptions} of the pseudo-view localizers from the axial localizer and filled the gap in {\color{red}the} existing literature.
Overall, we believe that this work has opened a new direction for automatic view planning of anatomy-oriented medical imaging beyond CMR.% based on the encouraging results

%{\color{yellow}\textbf{ToDo}: self-supervision (?); using the same terms consistently (R1); direct prescription of std. views from axial scout (R2); clinical usability (R4); identical figures and tables (R4).}

%{\color{yellow}Failure analysis.}

%\newpage     %This is set up for tables and figures at end of text but they
	     %may be in the text and is strongly preferred that way by some
	     %referees.

%\section{Tables}
%\begin{table}[htbp]
%\begin{center}
%\captionv{10}{Short title for list of figures - can be blank}{Note that
%tables and figures are to be in the text now, not at the end as required
%prior to 2014.
%Caption goes here.  It can be long.   It can be long.   It can be
%long.   It can be long.   It can be long.
%\label{tab_label}
%\vspace*{2ex}
%}
%\begin{tabular} {|l|c|c|c|c|}
%%\begin{edtable}{tabular} {|l|c|c|c|c|}
%\hline
%        &               \multicolumn{2}{c|}{}               &
%\multicolumn{2}{c|}{}        \\
%Chamber &\multicolumn{2}{c|}{4 V / S}&\multicolumn{2}{c|}{Monte Carlo}\\
%\cline{2-5}
%        & L mm          & $\Delta$ keV  & L mm          & $\Delta$ keV   \\
%\hline
%        &               &               &               &        \\
%3C      & 7.6           & 19.3          &  8.0          & 19.8           \\
%&&&&  \vspace{-2mm}\\
%3C      & 7.6           & 19.3          &  8.0          & 19.8           \\
%        &               &               &               &        \\
%\hline
%\end{tabular}
%%\end{edtable}
%\end{center}
%\end{table}

%        \multicolumn{ncol}{format}{text}
%       \cline{col1-col2}

%\clearpage

% following only if there is an appendix
\section*{Conflict of Interest}
\addcontentsline{toc}{section}{\numberline{}Conflict of Interest}
The authors have no relevant conflict of interest to disclose.

\section*{References}
\addcontentsline{toc}{section}{\numberline{}References}
\vspace*{-12mm}

% Following assumes you are using bibtex. However, for submission to the
% journal you MUST explicitly INCLUDE THE REFERENCES IN THE TEX FILE.
% In that case you need the following

% \begin{thebibliography}{10}
% insert the .bbl file generated by bibtex here
	%This will be a series of entries from your .bib file formatted
	%something like
	%\bibitem{Me09}
        %{I.~Meijsing, B.~W.~Raaymakers, A.~J.~E.~Raaijmakers \it et al.},
        %\newblock {Dosimetry for the MRI accelerator: the impact of a
	%magnetic field on the response of a Farmer NE2571 ionization chamber},
        %\newblock Phys. Med. Biol. {\bf 54}, 2993 -- 3002 (2009).

% \end{thebibliography}

% The following is when using bibtex and picks up the example.bib file

%\bibliography{Explicit address of .bib file}
%\bibliography{./refs}      %example.bib is on the same directory

\begin{thebibliography}{}

\end{thebibliography}


\begin{thebibliography}{10}

\bibitem{murphy2018mortality}
S.~L. Murphy, J.~Xu, K.~D. Kochanek, and E.~Arias,
\newblock {Mortality in the United States, 2017},
\newblock NCHS Data Brief {\bf 328} (2018).

\bibitem{vos2020global}
T.~Vos, S.~S. Lim, C.~Abbafati, et~al.,
\newblock Global burden of 369 diseases and injuries in 204 countries and
  territories, 1990--2019: a systematic analysis for the {Global Burden of
  Disease Study} 2019,
\newblock The Lancet {\bf 396}, 1204--1222 (2020).

\bibitem{la2012cardiac}
A.~La~Gerche, G.~Claessen, A.~Van~de Bruaene, et~al.,
\newblock Cardiac {MRI}: a new gold standard for ventricular volume
  quantification during high-intensity exercise.,
\newblock Circ. Cardiovasc. Imaging {\bf 6}, 329--338 (2012).

\bibitem{bai2019self}
W.~Bai, C.~Chen, G.~Tarroni, et~al.,
\newblock Self-supervised learning for cardiac {MR} image segmentation by
  anatomical position prediction,
\newblock in {\em MICCAI}, pages 541--549, Springer, 2019.

\bibitem{painchaud2019cardiac}
N.~Painchaud, Y.~Skandarani, T.~Judge, O.~Bernard, A.~Lalande, and P.-M.
  Jodoin,
\newblock Cardiac {MRI} segmentation with strong anatomical guarantees,
\newblock in {\em MICCAI}, pages 632--640, Springer, 2019.

\bibitem{robinson2017automatic}
R.~Robinson, V.~V. Valindria, W.~Bai, et~al.,
\newblock Automatic quality control of cardiac {MRI} segmentation in
  large-scale population imaging,
\newblock in {\em MICCAI}, pages 720--727, Springer, 2017.

\bibitem{wei2013comprehensive}
D.~Wei, Y.~Sun, S.-H. Ong, P.~Chai, L.~L. Teo, and A.~F. Low,
\newblock A comprehensive {3-D} framework for automatic quantification of late
  gadolinium enhanced cardiac magnetic resonance images,
\newblock IEEE. Trans. Biomed. Eng. {\bf 60}, 1499--1508 (2013).

\bibitem{zotti2018convolutional}
C.~Zotti, Z.~Luo, A.~Lalande, and P.-M. Jodoin,
\newblock Convolutional neural network with shape prior applied to cardiac
  {MRI} segmentation,
\newblock IEEE J. Biomed. Health Inform. {\bf 23}, 1119--1128 (2018).

\bibitem{chen2019learning}
C.~Chen, C.~Biffi, G.~Tarroni, S.~Petersen, W.~Bai, and D.~Rueckert,
\newblock Learning shape priors for robust cardiac {MR} segmentation from
  multi-view images,
\newblock in {\em MICCAI}, pages 523--531, Springer, 2019.

\bibitem{kramer2020standardized}
C.~M. Kramer, J.~Barkhausen, C.~Bucciarelli-Ducci, S.~D. Flamm, R.~J. Kim, and
  E.~Nagel,
\newblock Standardized cardiovascular magnetic resonance imaging ({CMR})
  protocols: 2020 update,
\newblock J. Cardiovasc. Magn. Reson. {\bf 22}, 1--18 (2020).

\bibitem{suinesiaputra2015quantification}
A.~Suinesiaputra, D.~A. Bluemke, B.~R. Cowan, et~al.,
\newblock Quantification of {LV} function and mass by cardiovascular magnetic
  resonance: multi-center variability and consensus contours,
\newblock J. Cardiovasc. Magn. Reson. {\bf 17}, 1--8 (2015).

\bibitem{lu2011automatic}
X.~Lu, M.-P. Jolly, B.~Georgescu, et~al.,
\newblock Automatic view planning for cardiac {MRI} acquisition,
\newblock in {\em MICCAI}, pages 479--486, Springer, 2011.

\bibitem{frick2011fully}
M.~Frick, I.~Paetsch, C.~den Harder, et~al.,
\newblock Fully automatic geometry planning for cardiac {MR} imaging and
  reproducibility of functional cardiac parameters,
\newblock J. Magn. Reson. Imaging {\bf 34}, 457--467 (2011).

\bibitem{alansary2018automatic}
A.~Alansary, L.~Le~Folgoc, G.~Vaillant, et~al.,
\newblock Automatic view planning with multi-scale deep reinforcement learning
  agents,
\newblock in {\em MICCAI}, pages 277--285, Springer, 2018.

\bibitem{blansit2019deep}
K.~Blansit, T.~Retson, E.~Masutani, N.~Bahrami, and A.~Hsiao,
\newblock Deep Learning-based Prescription of Cardiac {MRI} Planes,
\newblock Radiol. Artif. Intell. {\bf 1}, e180069 (2019).

\bibitem{jing2020self}
L.~Jing and Y.~Tian,
\newblock Self-Supervised Visual Feature Learning With Deep Neural Networks: A
  Survey,
\newblock IEEE Trans. Pattern Anal. Mach. Intell. {\bf 43}, 4037--4058 (2021).

\bibitem{duan2019centernet}
K.~Duan, S.~Bai, L.~Xie, H.~Qi, Q.~Huang, and Q.~Tian,
\newblock {CenterNet}: Keypoint triplets for object detection,
\newblock in {\em ICCV}, pages 6569--6578, 2019.

\bibitem{law2018cornernet}
H.~Law and J.~Deng,
\newblock {CornerNet}: Detecting objects as paired keypoints,
\newblock in {\em ECCV}, pages 734--750, 2018.

\bibitem{zhou2019bottom}
X.~Zhou, J.~Zhuo, and P.~Krahenbuhl,
\newblock Bottom-up object detection by grouping extreme and center points,
\newblock in {\em CVPR}, pages 850--859, 2019.

\bibitem{newell2016stacked}
A.~Newell, K.~Yang, and J.~Deng,
\newblock Stacked hourglass networks for human pose estimation,
\newblock in {\em ECCV}, pages 483--499, Springer, 2016.

\bibitem{wei2021training}
D.~Wei, K.~Ma, and Y.~Zheng,
\newblock Training Automatic View Planner for Cardiac {MR} Imaging via
  Self-Supervision by Spatial Relationship between Views,
\newblock in {\em MICCAI}, pages 526--536, Springer, 2021.

\bibitem{ginat2011cardiac}
D.~T. Ginat, M.~W. Fong, D.~J. Tuttle, S.~K. Hobbs, and R.~C. Vyas,
\newblock {Cardiac imaging: Part 1, MR pulse sequences, imaging planes, and
  basic anatomy},
\newblock Am. J. Roentgenol. {\bf 197}, 808--815 (2011).

\bibitem{10.1093/eurheartj/ehw680}
B.~Herzog,
\newblock {The CMR pocket guide App},
\newblock Eur. Heart J. {\bf 38}, 386--387 (2017).

\bibitem{pfister2015flowing}
T.~Pfister, J.~Charles, and A.~Zisserman,
\newblock Flowing {ConvNets} for human pose estimation in videos,
\newblock in {\em ICCV}, pages 1913--1921, 2015.

\bibitem{kingma2014adam}
D.~P. Kingma and J.~Ba,
\newblock Adam: A method for stochastic optimization,
\newblock arXiv preprint arXiv:1412.6980  (2014).

\bibitem{ronneberger2015u}
O.~Ronneberger, P.~Fischer, and T.~Brox,
\newblock {U-Net}: Convolutional networks for biomedical image segmentation,
\newblock in {\em MICCAI}, pages 234--241, Springer, 2015.

\bibitem{zeiler2014visualizing}
M.~D. Zeiler and R.~Fergus,
\newblock Visualizing and understanding convolutional networks,
\newblock in {\em ECCV}, pages 818--833, Springer, 2014.

\bibitem{danilouchkine2005operator}
M.~G. Danilouchkine, J.~J. Westenberg, A.~de~Roos, J.~H. Reiber, and B.~P.
  Lelieveldt,
\newblock Operator induced variability in cardiovascular {MR}: left ventricular
  measurements and their reproducibility,
\newblock J. Cardiovasc. Magn. Reson. {\bf 7}, 447--457 (2005).

\bibitem{kirillov2023segment}
A.~Kirillov, E.~Mintun, N.~Ravi, et~al.,
\newblock Segment anything,
\newblock arXiv preprint arXiv:2304.02643  (2023).

\bibitem{brown2020language}
T.~Brown, B.~Mann, N.~Ryder, et~al.,
\newblock Language models are few-shot learners,
\newblock Advances in neural information processing systems {\bf 33},
  1877--1901 (2020).

\end{thebibliography}
% above points to where we find the master reference list
% and also causes the bibliography to be printed

% When creating your bibliography you should run bibtex on your local
% computer after running pdflatex on your .tex file. bibtex will
% generate a .bbl file.
% Copy the contents of this .bbl file into your main latex document,
% replacing the "\bibliography" command which was pointing at your .bib file.

% following defines style of .bbl file

%\bibliographystyle{explicit relative path to medphy.bst}
\bibliographystyle{./medphy.bst}    %if this is installed on your system,
				    %it is not essential to have the    ./

% Note that you need to typeset once, then run bibtex, then typeset another
% two times to get the references working properly.

\end{document}